\documentclass[withindex,glossary]{cam-thesis}
\usepackage{url}
\usepackage{hyperref}
\usepackage[numbers]{natbib}
\usepackage{amsmath}
\usepackage{amssymb}
\usepackage{multirow}
\usepackage{verbatim}
\usepackage{float}
\usepackage{enumerate}
\newtheorem{theorem}{Theorem}

\newtheorem{definition}[theorem]{Definition}
\newcounter{examplecounter}
\newenvironment{example}{\begin{quote}%
    \refstepcounter{examplecounter}%
  \textbf{Example \arabic{examplecounter}}%
  \quad
}{%
\end{quote}%
}

\newtheorem{remark}[theorem]{Remark}

\usepackage[table]{xcolor}

\title{Entanglement Entropy and Algebra in Quantum Field Theory}

\author{Ahmed Halawani}

\college{Wolfson College}

\collegeshield{CollegeShields/Wolfson}

\submissiondate{May, 2023}

\submissionnotice{This essay is submitted for the degree of Master of Advanced Study in Mathematics}

\date{May, 2023}

\subjectline{Applied Mathematics}

\let\cleardoublepage\clearpage
\abstract{% 

Quantum Field Theory (QFT) represents a vast generalization of Quantum Mechanics (QM), as it deals with systems that have an infinite number of degrees of freedom. The Stone-von Neumann theorem, which establishes the equivalence of irreducible representations of the canonical commutation relations (CCR) in QM, does not extend to QFT. Consequently, QFT admits multiple inequivalent irreducible representations, leading to a much richer algebraic structure. This essay aims to explore the physics of QFT from the operator algebra perspective, particularly focusing on entanglement entropy. We discuss the role of von Neumann algebras of different types in QFT, describe the local operator algebra approach to QFT, and explain how entanglement entropy can be defined in terms of the algebra of observables. Additionally, we explore the benefits of this approach in concrete applications, specifically in  quantum field theory on curved spacetime.
}

\begin{document}

%%%%%%%%%%%%%%%%%%%%%%%%%%%%%%%%%%%%%%%%%%%%%%%%%%%%%%%%%%%%%%%%%%%%%%%%%%%%%%%%
%% Title page, abstract, declaration etc.:
%% -    the title page (is automatically omitted in the technical report mode).

\frontmatter{}

%%%%%%%%%%%%%%%%%%%%%%%%%%%%%%%%%%%%%%%%%%%%%%%%%%%%%%%%%%%%%%%%%%%%%%%%%%%%%%%%
%% Thesis body:
%%
\part{}
\chapter{Motivation}

One of the major goals of theoretical physics is to develop a unified description of all fundamental interactions, including gravity. Quantum Field Thoery (QFT) provides a framework for describing the interactions of elementary particles in terms of quantum fields. QFT represents a vast generalization of Quantum Mechanics (QM), as it deals with systems that have an infinite number of degrees of freedom. While QFT shares many similarities with QM, it also exhibits some important differences. In QM, the Stone-von Neumann theorem establishes the equivalence of irreducible representations of the canonical commutation relations. However, this theorem does not extend to QFT, where multiple inequivalent irreducible representations are possible, leading to a much richer algebraic structure, where different representations can describe different physical phenomena or states of the system. For example, in the context of QFT on a curved spacetime, there are different possible representations of the algebra of observables that can describe different gravitational effects, such as the bending of light or the redshift of light from distant sources. These different representations can correspond to different physical states of the system, leading to a more complex and diverse range of physical phenomena than is possible in quantum mechanics.

In this essay, we explore the difference between different formalisms of describing physical systems. We explain the physics of QFT from the operator algebra perspective, with a particular focus on entanglement entropy. We discuss the role of von Neumann algebras of different types in QFT, describe the local operator algebra approach to QFT, and explain how entanglement entropy can be defined in terms of the algebra of observables. Additionally, we explore the benefits of this approach in concrete applications, specifically in quantum field theory on curved spacetime. 

This essay is divided into two parts. Part I discusses the algebraic formalism in general. As we move on to Part II, we will relate everything discussed to QFT. It's meant to be self-contained, so all relevant background material of Part II will be discussed in Part I. Chapter \ref{2} starts by providing the background for the `standard' representation formalism and demonstrating its shortcomings, then it sets the ground for the tools needed for the algebraic formalism. Chapter \ref{3} discusses the algebraic formalism. It sheds light on the mathematical reasoning which underlies the operator algebra approach, and the advantages it has over the standard one. In Chapter \ref{4}, we apply this approach in QFT, see the advantages it holds, and how it plays a role in entanglement entropy. Finally, Chapter \ref{5} discusses more applications of the algebraic approach in QFT on curved spacetime, demonstrating that it can be more beneficial to consider different formalisms at times.

\chapter{Mathematical background}\label{2}

\section{Banach Space} \label{banach}

 Hilbert space underlies quantum mechanics and such a space is a special case of a \emph{Banach space}. We are interested in operators on Hilbert space, so first we should understand Banach space and operators on Banach space.
\begin{definition}
A \textit{Banach space} space is a  vector space $V$ which is complete with respect to a  norm:
\begin{equation}
     \|\cdot \|: V \rightarrow \mathbb{R}
\end{equation}
which satisfies:
\begin{align}
    &(i)\quad \|f\| \geq 0 &&\text{(non-negativity)}\\
    &(ii) \quad \|f\|=0 \Leftrightarrow f=0 &&\text{(definiteness)}\\
    &(iii) \quad \|\lambda  \cdot f\|=|\lambda |\|f\| &&\text{(homogeneity)}\\
    &(iv) \quad  \|f+g\| \leq\|f\|+\|g\| &&\text{(sub-additivity)}
\end{align}
$\forall \lambda \in \mathbb{C}$ and $f,g \in V$
\end{definition}

\subsection{Bounded Operators}

A standard topic in mathematics is the study of a structure through the study of maps between different instances of the structure. In the case of Banach space, we can begin by studying a linear map that starts in a set and goes to a different set with more structure:
\begin{equation}
    A: V \rightarrow W,
\end{equation}
where $V$ ($V, \| \cdot \|_V$ ) is a general set with only normed space, and the target set $W$ is a Banach space ($W, \| \cdot \|_W$ ), which has a complete norm, i.e. more structure. The map $A$ is \textit{bounded} if:
\begin{equation} 
    \sup _{f \in V} \frac{\|A f\|_W}{\|f\|_V}<\infty.
\end{equation}
By scaling $f$ to a unit vector, this statement can be equivalently states as:
\begin{equation}
    \sup _{{\|f\|_V =1}}  \; \|A f\|_W <\infty
\end{equation}
and this defines the \textit{operator norm} $\|A\|$ of a bounded operator. 
\begin{definition}
    A Banach algebra $\mathcal{A}$ is a Banach space (complete with respect to its norm) such that $\|AB\|\leq\|A\|\|B\|$ for all $A,B\in\mathcal{A}$. The algebra of bounded operators on a Banach space is a Banach algebra.
\end{definition}

Unbounded operators, on the other hand, have no such norm. Bounded operators are of importance because often unbounded operators are reduced to a sequence of bounded ones \cite{Segal1947IrreducibleRO}. Furthermore, because they are bounded, we can write the commutation relation between them.

\begin{example}
Take $V=W$. Consider the identity map on $W$, $id_W: W \rightarrow W$. Then
\begin{equation}
    \|id_W\| = \sup _{{\|f\|_W }=1}  \|id_W f \|_W =  \sup _{{\|f\|}=1}  \|f\| = 1 < \infty
\end{equation}
Hence the identity map is bounded and the operator norm of the identity map is 1.
\end{example}

\section{Hilbert Spaces}
 A Hilbert space $\mathcal{H}$ is a standard too used to represent the state space of a quantum system. A Hilbert space is a type of vector space that is equipped with an inner product that is Cauchy \textit{complete} \cite{kennedy2013hilbert}. A metric space $(V, d)$ is called complete when every Cauchy sequence\footnote{ A sequence $\left(f_n\right)$ is a Cauchy sequence in $V$ when $\left\|f_n-f_m\right\| \rightarrow 0$ when $n, m \rightarrow \infty ;$ more precisely, for any $\varepsilon>0$ there is $N \in \mathbb{N}$ such that $\left\|f_n-f_m\right\|<\varepsilon$ for all $n, m>N$. A sequence $\left(f_n\right)$ converges if there is $f \in V$ such that $\lim _{n \rightarrow \infty}\left\|f_n-f\right\|=0$. \cite{landsman1998lecture}} converges. This inner product defines a norm.\\

\begin{definition}
An \textit{inner product} on a complex linear space X is a map
\begin{equation}
    \langle \; . \; | \; . \;\rangle: X \times X \rightarrow \mathbb{C}
\end{equation}
Such that, $\forall \;x,y,z \in X$ and $\lambda, \mu \in \mathbb{C}$:
    \begin{align*}
        &(a) \quad \langle x| \lambda y + \mu z\rangle = \lambda \langle x|y\rangle + \mu \langle x|z\rangle \quad && \text{(linearity in second argument)}\\
        &(b) \quad \langle y|x\rangle = \overline{\langle x|y\rangle} \quad && \text{(Hermitian Symmetry)}\\
        &(c) \quad \langle x|x\rangle \geq 0  \quad && \text{(non-negativity)}\\
        &(d) \quad \langle x|x\rangle = 0 \: \text{iff x = 0}  \quad && \text{(positive definiteness)}
    \end{align*}
    where $\overline{ \langle \; . \; | \; . \;\rangle}$ is complex conjugation.
\end{definition}
It follows from (a) and (b) that the inner product is anti-linear in the first term:
\begin{equation*}
    \langle\lambda x +\mu  y | z\rangle = \Bar{\lambda} \langle x|z\rangle + \Bar{\mu} \langle y|z\rangle.
\end{equation*}
The inner product $\langle \; \cdot \; | \; \cdot \;\rangle$ induces a  norm $||\cdot||$ via the relation:
\begin{equation}
    ||x|| = \sqrt{\langle x|x\rangle}
\end{equation}
and closure with respect to this norm gives us a \emph{Hilbert space} $\mathcal{H}$.
\begin{definition}
    A Hilbert space is called separable if it admits a countable complete orthonormal basis.
\end{definition}

\subsection{Operators on Hilbert space: $C^\ast$-algebra}
$C^*$-algebras play a crucial role in the algebraic approach to quantum mechanics, as they provide a rigorous mathematical framework that unifies various aspects of the theory. Their importance lies in their ability to naturally model the algebra of bounded linear operators on a Hilbert space, capture the properties of quantum observables and states, and facilitate the development of powerful mathematical tools for analyzing quantum systems. Observables, being self-adjoint operators, naturally form a $C^*$-algebra.

\begin{definition}
    A Banach $\ast$-algebra $\mathcal{A}$  is a Banach algebra that is also closed under involution $\ast:\mathcal{A}\rightarrow\mathcal{A}$, where $(a^*)^* = a$, and is  antihomomorphism: $(ab)^* = b^*a^*$  $\forall a, b \in  A$. Further, $\mathcal{A}$ is a $C^\ast$-algebra if $\|A^\ast A\|=\|A\|^2$.
\end{definition}
We can also think of a compatibility condition between the norm and involution:
\begin{equation}
     \|a^*\| = \|a\| \quad \quad  \forall a,\in \text{A}.
 \end{equation}
\begin{remark}
 Every $C^\ast$-algebra may be represented as a sub-algbera of the algebra of bounded operators $\mathcal{B}(\mathcal{H})$ on a Hilbert space $\mathcal{H}$.
\end{remark}

\section{Projectors and spectral theory}\label{proj}

Projectors in $\mathcal{B}(\mathcal{H})$ play a crucial role in the classification of algebra, as we will see their use in section (\ref{vn}). In essence, they form a basis for an algebra \cite{helmberg2008introduction}.
\begin{definition}
    A projector is a positive operator $P\in\mathcal{B}(\mathcal{H})$, which means it has the form $P=U^*U$. Further, it satisfies the \emph{idempotent} condition $P^2=P$.
\end{definition}
\begin{definition}
    A partial isometry $U\in\mathcal{B}(\mathcal{H})$ is such that $U^\ast U=E$ and $UU^\ast=F$ where both $E$ and $F$ are projectors. In particular, $U$ is an isometry if $E=I$, a co-isometry if $F=I$ and unitary if $E=I=F$, where $I$ is the identity operator \cite{muger}. If two projectors $E$ and $F$ are related by a partial isometry in this way we say that they are equivalent and write $E\sim F$.
\end{definition}
We can see how a linear transformation leads to definitions of rank-1 projectors by defining $U = |\phi\rangle\langle\psi|, E = |\phi\rangle\langle\phi|$, and $F = |\psi\rangle\langle\psi|$:

\begin{equation}
    UU^* = |\phi\rangle\langle\psi|\psi\rangle\langle\phi| = |\phi\rangle\langle\phi| = E
\end{equation}
and
\begin{equation}
    UU^* = |\psi\rangle\langle\phi|\phi\rangle\langle\psi| = |\psi\rangle \langle\psi| = F.
\end{equation}
\begin{definition}
   A self-adjoint operator satisfies $x=x^\ast$. Such an operator may be written as the difference of two positive operators. The \textit{spectral decomposition} of a self-adjoint operator $x$ is given by:
\begin{equation}
    x = \sum x_i E_i
\end{equation}
when the spectrum of $x$, $\{x_i\}\subset\mathbb{R}$ is countable. The operators $E_i$ are pairwise orthogonal projectors. 

The spectral decomposition for operators with continuous eigenbasis:
\begin{equation}
    x = \int x \; d E(x) \quad \quad (x \in \mathbb{R})
\end{equation}
where $dE(x)$ is a not a projector, but the integrals $\int_\Delta dE(x)$, where $\Delta\subset\mathbb{R}$ have non-zero measure with respect to $dx$, are projectors.
\end{definition}

\begin{example} Observables are given by self-adjoint operators. Say we have an observable  $x$ and we observe the value $x_k$ and infer that the system is subsequently in the state $|\phi_k\rangle$ which is an eigenvector of $x$ and defines the projection valued measure (PVM) $E_i =|\phi_i\rangle\langle\phi_i|$, then:
\begin{align}
    x |\phi_k\rangle &= \sum x_i |\phi_i\rangle\langle\phi_i|\phi_k\rangle= x_k |\phi_k\rangle,
\end{align}
where observation of the eigenvalue $x_k$ is believed to result in the collapse of the wavefunction $\phi\mapsto E_i \phi/\|E_i\phi\|=\phi_i$.
\end{example} 

A positive operator $P$ is also self-adjoint, and the spectrum of a positive operator is non-negative: $spec(P)\geq 0$. Notice that if $P$ is a projector we have $spec{P}\subseteq\{0,1\}$. This spectral positivity of such positive operators allows us to order operators as $E\leq Q$, which means that $Q-E\geq 0$, which means that $Q-E$ is a positive operator.

To describe algebras it is useful to introduce a preorder $\preceq$. For two projectors we say that $P\preceq Q$ if $P\sim E\leq Q$. We explain this by example: let $\mathcal{A} = \mathcal{M}_3 $, and let projectors $P,Q\in\mathcal{A}$ with $P\preceq Q$. such as:
\begin{equation}
    Q=\left[\begin{array}{lll}
1 & 0 & 0 \\
0 & 1 & 0 \\
0 & 0 & 0
\end{array}\right]=\left[\begin{array}{lll}
1 & 0 & 0 \\
0 & 0 & 0 \\
0 & 0 & 0
\end{array}\right]+\left[\begin{array}{lll}
0 & 0 & 0 \\
0 & 1 & 0 \\
0 & 0 & 0
\end{array}\right]
\end{equation}
so we call $Q$ a \textit{rank 2} projector. On the other hand consider:
\begin{equation}
    P=\frac{1}{3}\left[\begin{array}{lll}
1 & 1 & 1 \\
1 & 1 & 1 \\
1 & 1 & 1
\end{array}\right] = |\xi\rangle\langle\xi|
\end{equation}
with
\begin{equation}
    |\xi\rangle=\frac{1}{\sqrt{3}}\left[\begin{array}{l}
1 \\
1 \\
1
\end{array}\right]
\end{equation}
so  $P$ is a \textit{rank 1} projector. Hence $P \preceq Q$, \footnote{the symbol $\preceq$ reads $P$ is equivalently less than $Q$, which means $P$ has a fewer number of projections. This is not quite like ``less than".} which means that there is a projector $E$ that is of the same rank as $P$ and write $P\sim E$, such that $E\preceq  Q$
and that implies  ${0}\leq Q-E$, which means that $Q-E$ is a positive operator with non-negative eignevalues. $E$ can be defined as:
    \begin{equation}
        E=\left[\begin{array}{lll}
1 & 0 & 0 \\
0 & 0 & 0 \\
0 & 0 & 0
\end{array}\right]
    \end{equation}
because both of them are of rank 1, and it can be shown that they are related through the partial isometry $U$:
\begin{equation}
    U=\frac{1}{\sqrt{3}}\left[\begin{array}{lll}
1 & 0 & 0 \\
1 & 0 & 0 \\
1 & 0 & 0
\end{array}\right]
\end{equation}
where $P= UU^*$ and $E= U^* U$. Both of these projectors are examples of minimal \cite{sunder} projections, which is defined as follows. A \textit{minimal projector} $E\in \mathcal{A}$ is one that can be expressed as 
\begin{equation}
    E \mathcal{A} E=\mathbb{C} E
\end{equation}
\begin{example}
    Let $\mathcal{A} = \mathcal{M}_3 $, the $E$, as given above, is a minimal projector:
    \begin{equation}
        E \mathcal{A} E = \left[\begin{array}{lll}
1 & 0 & 0 \\
0 & 0 & 0 \\
0 & 0 & 0
\end{array}\right]\left[\begin{array}{lll}
* & * & * \\
* & * & * \\
* & * & *
\end{array}\right]\left[\begin{array}{lll}
1 & 0 & 0 \\
0 & 0 & 0 \\
0 & 0 & 0
\end{array}\right] = \left[\begin{array}{lll}
* & 0 & 0 \\
0 & 0 & 0 \\
0 & 0 & 0
\end{array}\right]=\mathbb{C} E
    \end{equation}
where $*$ is an arbitrary complex number.
\end{example}
\begin{remark}
Note that if we introduce $R \sim E \sim P$, such that:
\begin{equation}
    R=\left[\begin{array}{lll}
0 & 0 & 0 \\
0 & 0 & 0\\
0 & 0 & 1
\end{array}\right]
\end{equation}
then $Q-R \ngeq {0}$, because it contains a negative eigenvalue. So we see that $Q \ngeq R$, but  $Q \succeq R$.
    \end{remark}

\subsection{Quantum Mechanics in Hilbert Space}

A quantum system is defined on an associated complex Hilbert space $\mathcal{H}$.
We can assume that the \textit{states} of the system are all the positive, normalised,  linear maps $\omega : \mathcal{A} \rightarrow \mathbb{C}$, where $\mathcal{A}\subseteq \mathcal{B}(\mathcal{H})$. States may be given by a density operator $\varrho$ which is a positive, trace-class operator.
\begin{definition}
     An operator $T$ is said to be \textit{trace class} if for $T^\dag T := \varrho^2$, with $\varrho^\dag = \varrho=|T|$, we have $\operatorname{Tr}[\varrho] < \infty$.
     \end{definition}
That is, 
\begin{equation}
    \omega(a):=\operatorname{Tr}[\varrho a],
\end{equation}
and a state $\omega$ is called \textit{pure} if $\exists$ normalised  $\psi \in \mathcal{H}$ such that:
\begin{equation}
    \omega(a)={\langle\psi|a|\psi\rangle}, \quad \quad \forall \;a\in \mathcal{A},
\end{equation}
such that $\varrho=|\psi\rangle\langle\psi|$. As a result, for each pure state $\omega$, we can connect it to an element $\psi$ within the Hilbert space. However, this association does not create a unique, one-to-one relationship between the two. This lack of a direct one-to-one correspondence is essential to consider when working with the algebraic approach to quantum mechanics and understanding the underlying structure of quantum systems.

\subsubsection{Expectation Value in terms of $\varrho$} \label{Expectation Value in terms of}

We have motivated the discussion on the density matrix by mentioning that is used widely in physics. In particular, it can be used to describe states in a system, and that it offers a way to calculate the expected value. The expectation value is the probability of an outcome of observable  $O$ when measured in state $\psi$, and it's defined per Born rule:

\begin{equation} \label{exp value}
   \langle O \rangle_\psi = \langle \psi | O  | \psi\rangle 
\end{equation}
Where the lower index represents the state. In this example, the state $\psi$ can be represented in basis in $\mathcal{H}$ as:

\begin{equation} \label{exp basis}
     | \psi\rangle  = \sum_n a_n | \phi_n \rangle
\end{equation}
Putting equations \ref{exp value} and \ref{exp basis} together, we can express the expectation value as:

\begin{equation}
    \langle O \rangle_\psi = \sum_n  p_n  \langle \phi_n |O| \phi_n \rangle
\end{equation}
Where $p_n = |a_n|^2 = a_n ^* a_n$. Also, for any operator $O \in \mathcal{A}$ we can express this expectation value in terms of the density matrix a follows:

\begin{equation}
      \langle O \rangle_\psi =  \operatorname{Tr}[\varrho_\psi \; O] = \operatorname{Tr}[| \psi \rangle \langle \psi | \; O]
\end{equation}
This is the first use of the density matrix we have encountered in this essay thus far. We shall later see how it plays a viable role in representing entropy.

Quantum mechanics makes use of the mathematical structure of Hilbert space to describe the state space of a quantum system. The set of all possible wave functions of a quantum system is called the state space of the system. The state space of a quantum system is a Hilbert space, and the wave function of the system is a vector in that Hilbert space. The mathematical structure of the Hilbert space allows us to use the tools of linear algebra, such as vector addition and scalar multiplication, to describe the state of a quantum system and the transformations between different states. The inner product of two vectors in the Hilbert space is used to calculate the probability of finding the system in a particular state, and the norm of a vector in the Hilbert space is used to define the concept of distance between two wave functions. More generally, the probability that a \textit{measurement} \cite{schuller} of an observable $A$ on a system that is in the state $\omega$ yields a result in the Borel set $E \subseteq \mathbb{R}$ is given by:
\begin{equation}
\mu_\omega^A(E):=\operatorname{Tr}\left[\mathrm{P}_A(E)\varrho\right],
\end{equation}
Where the map  $\mathrm{P}_A: \operatorname{Borel}(\mathbb{R}) \rightarrow \mathcal{L}(\mathcal{H})$ serves as a special connection between Borel-measurable subsets of the real numbers and Banach space of bounded linear maps on the Hilbert space, and it is uniquely determined by the self-adjoint map $A$ according to the principles of the spectral theorem \cite{santi}. More on this will follow in section (\ref{banach}).

\subsection{Density Matrix $\varrho$} \label{density}

The density matrix is a powerful tool in physics. As we have seen in section \ref{Expectation Value in terms of}, it allows us to describe systems that are in pure and mixed states, and it provides a way of calculating the expected values of observables for systems \cite{witten2020mini}. For example, a quantum field could be in a state where it has some particles with a certain energy and some `particles' with another energy. It provides a way of encoding this information about the distribution of particles over different states and it provides a way of calculating the expectation values of observables for the field. We can define the density matrix for a state $\omega$ on $\mathcal{M}^2$, as $\rho  = R^\dag R\geq\rho^2$, such that for $a\in\mathcal{M}_2$ we have $\omega(a)=\operatorname{Tr}[\rho a]$. We have 
\begin{equation}
    R =\begin{pmatrix}
            \sqrt{p_1} \langle \psi_1 | \\
            \sqrt{p_2}  \langle \psi_2 |  \\
            \end{pmatrix} ,
\end{equation}\\
where the probability weight $0 \leq p_i \leq 1$ and $\sum_i p_i =1$, and $|\psi_i\rangle$ lives in the Hilbert space $\mathbb{C}^2$.
\begin{itemize}
    \item In the case $\rho^2 = \rho : \rho$ is a projector  $| \psi \rangle \langle \psi |$ corresponding to a pure state. A state is pure when $p_1= 1$ and $p_2 =0$, or the opposite.
    \item In the case $\rho^2 < \rho : \rho= p_1 | \psi_1 \rangle \langle \psi_1 | + p_2 | \psi_2 \rangle \langle \psi_2 |$ corresponding to a mixed state. A \textit{maximally mixed} state is when $p_1= p_2 = \frac{1}{2}$. 
\end{itemize}

 We can define a \textit{separable density matrix} for a 2-qubit system given by the Hilbert space $\mathbb{C}^2\otimes\mathbb{C}^2=\mathbb{C}^4\ni \Psi_i=\psi_i\otimes\phi_i$, as
\begin{equation} \label{density eq}
    \varrho = \sum_i p_i | \psi_i \rangle \langle \psi_i|\otimes | \phi_i \rangle \langle \phi_i|
\end{equation}
We see that $\varrho$ has the form of a sum between distinct separable pure states $|\psi_i\rangle\langle\psi_i|\otimes|\phi_i\rangle\langle\phi_i|$. A state that is not separable is called \emph{entangled}.

\begin{example}
A separable mixed density matrix of 2 qubits given in the $Z$-basis (or standard basis) $| \psi_1 \rangle= | \phi_1 \rangle =
\begin{pmatrix}
            1 \\
            0 \\
            \end{pmatrix} $ and $| \psi_2 \rangle =  | \phi_2 \rangle =\begin{pmatrix}
            0 \\
            1 \\
            \end{pmatrix} $ is given by:

            \begin{align}
                \varrho &= \sum^2_{i=1} p_i (| \psi_i \rangle \langle \psi_i| \otimes | \phi_i \rangle \langle \phi_i |)\\
                & = p_1 (| \psi_1 \rangle \langle \psi_1| \otimes | \phi_1 \rangle \langle \phi_1 |) + p_2 (| \psi_2 \rangle \langle \psi_2| \otimes | \phi_2 \rangle \langle \phi_2 |)\\
                &= p_1 \begin{pmatrix}
            1 & 0 \\
            0 & 0\\
            \end{pmatrix} \otimes \begin{pmatrix}
            1 & 0 \\
            0 & 0\\
            \end{pmatrix} + p_2 \begin{pmatrix}
            0 & 0 \\
            0 & 1\\
            \end{pmatrix} \otimes \begin{pmatrix}
            0 & 0 \\
            0 & 1\\
            \end{pmatrix}\\
            &= \begin{pmatrix}
            p_1 & 0 & 0 & 0 \\
            0 & 0 & 0 & 0\\
            0 & 0 & 0 & 0\\
            0 & 0 & 0 & p_2\\
            \end{pmatrix}
            \end{align} 
In the second line we clearly see that this density state is expressed in terms of a sum of separable pure states, thus it's mixed.
\end{example}
Otherwise, $\Psi\in\mathbb{C}^2\otimes\mathbb{C}^2$ would be in  $\Psi=\sum c_i\Psi_i$. We can define a general \textit{entangled pure states density matrix} as:
\begin{equation} 
    \varrho = \sum_{i,j} c_i \Bar{c_j}| \Psi_i \rangle \langle \Psi_j|
\end{equation}
\begin{example}
 An entangled state of 2 qubits given in the $Z$-basis (or standard basis) $| \psi_1 \rangle= | \phi_1 \rangle =
\begin{pmatrix}
            1 \\
            0 \\
            \end{pmatrix} $ and $| \psi_2 \rangle =  | \phi_2 \rangle =\begin{pmatrix}
            0 \\
            1 \\
            \end{pmatrix} $, where $|\Psi\rangle = \sum c_i |\psi_i\rangle\otimes|\phi_i\rangle$ is given by
    \begin{align}\label{rho exmp}
        \varrho & = |\Psi \rangle \langle \Psi| \\
        &= \left(\sum_{i} c_i | \psi_i \rangle \otimes | \phi_i \rangle \right)\left(\sum_{j}\Bar{c_j}\langle \psi_j|\otimes \langle\phi_j |\right)= \sum_{i,j} c_i \Bar{c_j}(| \psi_i \rangle \langle \psi_j| \otimes | \phi_i \rangle \langle \phi_j |)\\
         & = c_1 \Bar{c_1}(| \psi_1 \rangle \langle \psi_1|\otimes | \phi_1 \rangle \langle \phi_1 |) + c_2 \Bar{c_2}(| \psi_2 \rangle \langle \psi_2|\otimes |\phi_2 \rangle \langle \phi_2|)\\
         &+ c_1 \Bar{c_2}(| \psi_1 \rangle \langle \psi_2| \otimes | \phi_1 \rangle \langle \phi_2 |) + c_2 \Bar{c_1}(| \psi_2 \rangle \langle \psi_1|\otimes | \phi_2 \rangle \langle \phi_1 |)\\
         &= |c_1|^2 \begin{pmatrix}
            1 & 0 \\
            0 & 0\\
            \end{pmatrix} \otimes \begin{pmatrix}
            1 & 0 \\
            0 & 0\\
            \end{pmatrix} + |c_2|^2 \begin{pmatrix}
            0 & 0 \\
            0 & 1\\
            \end{pmatrix} \otimes \begin{pmatrix}
            0 & 0 \\
            0 & 1\\
            \end{pmatrix}\\
            &+ \ c_1 \Bar{c_2} \begin{pmatrix}
            0 & 1 \\
            0 & 0\\
            \end{pmatrix} \otimes \begin{pmatrix}
            0 & 1 \\
            0 & 0\\
            \end{pmatrix} + c_2 \Bar{c_1} \begin{pmatrix}
            0 & 0 \\
            1 & 0\\
            \end{pmatrix} \otimes \begin{pmatrix}
            0 & 0 \\
            1 & 0\\
            \end{pmatrix}\\
            &= \begin{pmatrix}
            p_1 & 0 & 0 & c_1 \Bar{c_2} \\
            0 & 0 & 0 & 0\\
            0 & 0 & 0 & 0\\
            c_2 \Bar{c_1} & 0 & 0 & p_2\\ 
            \end{pmatrix} \quad \quad (\text{\textit{pure}})
    \end{align} 

In the third line we have the interference terms. To arrive at the final matrix we used the substitution $|c_1|^2 = p_1$ and $|c_2|^2 =p_2$. Since the outcome can be expressed in terms of the outer product of states, the density matrix is pure and entangled.
\end{example}
\textit{Entanglement} refers to the property of a composite quantum system consisting of two or more subsystems that cannot be described independently of each other. If the density matrix of the composite system cannot be written as a tensor product of the density matrices of the individual subsystems, then the subsystems are said to be entangled\cite{witten2020mini}.

\subsection{Observables \& Representations in Hilbert Space}

Similar to how one vector can be expressed in different bases and coordinate systems, such as Cartesian or spherical coordinates, a state-vector $| \psi \rangle$ is a member of $\mathcal{H}$. While vectors are invariant with respect to the chosen basis, their components are covariant.

The observables of a quantum system are represented by operators, which are linear transformations on the state space of the system. Observables are physical quantities that can be measured, such as position, momentum, energy, and entropy. A representation of a physical system is a way of describing the state space of the system and the observables of the system using a specific mathematical structure. In the case of Hilbert space, a representation is a way of describing the state space and observables of a physical system using a specific Hilbert space, along with a set of operators that correspond to the observables of the system. Operators are linear transformations on the state space of the system. For all  $| \psi \rangle \, , | \phi \rangle \, \in \mathcal{H}$ and $\langle \psi | \, , \langle \phi | \, \in \mathcal{H^*}$, an operator $\Hat{A}$ corresponds to an observable A if and only if it satisfies the following conditions:
    \begin{align*}
        &(i) \quad \Hat{A} \; (a \, |\psi\rangle \, + b \, | \psi \rangle) = \,  a\, \Hat{A}|\psi\rangle + b \, \Hat{A} | \psi \rangle && \text{(linearity.} \;  \forall a,b \in \mathbb{C})\\
        &(ii) \quad \langle \psi | \Hat{A} | \psi \rangle \, \in \mathbb{C} && \text{(Hermicity)}\\
        &(iii) \quad \langle \psi | \Hat{A} | \phi \rangle  = \overline{\langle \phi | \Hat{A} | \psi \rangle} \quad && \text{(self-adjointness)}
    \end{align*}
There are unaccountably many  representations in Hilbert space, but the most common ones are the position representation and the momentum representation.  The Stone-von Neumann, which will be discussed in section (\ref{s vn}),  states that any two irreducible representations of the canonical commutation relation are unitarily equivalent, meaning that they are essentially the same up to a unitary transformation In essence, $[\hat{x},\hat{p}]$ yields the same value regardless of the space they are represented on.

The \textit{position representation} is a representation of a quantum system in which the state space of the system is represented by a Hilbert space of functions of position. In this representation, the wavefunction of a system is a function of position, denoted by $\psi(x)$, and the position operator, denoted by $\hat{x}$. The action of the position operator on a wave function is simply multiplication by the position coordinate:
\begin{equation} \label{prep1}
    \hat{x} \, |\psi(x) \rangle = x \, |\psi(x)\rangle,
\end{equation}
Where  $\hat{x}$ is the position operator, $x$ is the eigenvalue corresponding to  the eigenstate $|\psi(x)\rangle$.

The momentum operator, denoted by $\hat{p}$, is represented by the derivative operator. The action of the momentum operator on a wave function is given by the derivative of the wave function with respect to position:
\begin{equation} {\label{prep2}}
    \hat{p} \, |\psi(x) \rangle = -i \hbar \, \frac{d}{dx} |\psi(x)\rangle,
\end{equation}
where  $\hat{p}=-i \hbar \, \frac{d}{dx}$ is the momentum operator and its eigenvalues are given on momentum eigenstates in the momentum representation.

The \textit{momentum representation} is a representation of a quantum system in which the state space of the system is represented by a Hilbert space of functions of momentum. In this representation, the wave function of a system is a function of momentum, denoted by $\phi (p)$, and the momentum operator, denoted by $\hat{p}$. The action of the momentum operator on a wave function is given by the derivative of the wave function with respect to position:
\begin{equation} {\label{mrep1}}
    \hat{p} \, |\phi(p)\rangle = p \, |\phi(p)\rangle,
\end{equation}
where  $\hat{p}$ is the momentum operator, $p$ is the eigenvalue corresponding to the momentum operator, and $\phi(p)$ is the complex eigenstate.
The position operator is represented by the derivative operator here. The action of the position operator on a wavefunction is given by the derivative of the wavefunction with respect to the momentum:
\begin{equation} {\label{mrep2}}
    \hat{x} \, |\phi(p) \rangle = i \hbar \, \frac{d}{dp} |\phi(p)\rangle.
\end{equation}
We shall see in section (\ref{Fourier link}) that there's a link between $\psi(x)$ and $\phi(p)$, as they stem from the same vector.

\section{Uniqueness of Representations}

As quantum mechanics was developing, it was important to ask:
\emph{Can one begin with a classical theory, apply the quantization principles, and ultimately obtain two distinct quantized theories? }\cite{stepanian2017unitary}
This question was answered by Marshall Stone and John von Neumann. Their findings' direct implication is that all quantized theories derived through canonical quantization from a classical theory are physically indistinguishable from one another. In the following subsection, we will delve deeper into the implications of this result.

\subsection{Stone's Theorem}

Stone's theorem shows there is a one-to-one correspondence between self-adjoint operators and strongly continuous one-parameter unitary groups  \cite{stone1930linear}. We can motivate Stone's theorem by asking the following two questions:
\begin{enumerate}[i]
    \item  How  are observables constructed?
    \item  Given that $U(t) \circ U(s) = U(t+s)$, and $U(0) = id_\mathcal{H}$, how arbitrary is the stipulation that the unitary dynamics is controlled by  $U(t) = e^{-itH}$, for some self-adjoint operator $H$? 
\end{enumerate}
Skipping ahead, the answers to both of these questions are provided by Stone's theorem: (1) obserables come naturally from the generators of a group. (2) It it's not arbitrary at all- it's fixed. We can show this by considering
 a one-parameter, unitary, strongly continuous group:
\begin{equation}
    G := \{U(t): \mathcal{H} \rightarrow \mathcal{H} |\; t\in \mathbb{R}, \; U^*(t) U(t) = id_\mathcal{H} \}
\end{equation}
and the strong continuation condition:
\begin{equation}
    \forall \;\psi \in \mathcal{H}: \lim_{t \rightarrow t_0} (U(t) \psi) = U(t_0) \psi.
\end{equation}
Now we can define a unitary, Abelian, one-parameter, strongly continuous group  (UAOPG) $U(\cdot): \mathbb{R} \rightarrow \mathcal{L}(\mathcal{H})$ which perseves the norm of any $\psi$ in the Hilbert space. The generators of this group would be:
\begin{equation*}
    A: \mathcal{D}^{Stone}_A \rightarrow \mathcal{H}
\end{equation*}
where
\begin{equation} \label{Dstone}
    \mathcal{D}_A^{\text {Stone }}:=\left\{\psi \in \mathcal{H} \mid \lim _{\varepsilon \rightarrow 0} \frac{i}{\varepsilon}(U(\varepsilon) \psi-\psi) \text { exists }\right\}
\end{equation}
such that
\begin{equation}
    \psi \mapsto A \psi:=\lim _{\varepsilon \rightarrow 0} \frac{i}{\varepsilon}(U(\varepsilon) \psi-\psi).
\end{equation}
Note that this object is equivalent to taking $U(\cdot)$ and applying it to $\psi$, so we get a map $U(\cdot) \psi: \mathbb{R} \rightarrow \mathcal{H}$, then we take the derivative of this object and the evaluating it at zero. We can express this as:
\begin{equation}
    \lim _{\varepsilon \rightarrow 0} \frac{i}{\varepsilon}(U(\varepsilon) \psi-\psi)=i \lim _{\varepsilon \rightarrow 0} \frac{U(0+\varepsilon) \psi-U(0) \psi}{\varepsilon}:=i[U(\cdot) \psi]^{\prime}(0).
\end{equation}
So we can express (\ref{Dstone}) more compactly as:
\begin{equation}
    \mathcal{D}_A^S \equiv \mathcal{D}_A^{\text {Stone }}:=\left\{\psi \in \mathcal{H} \mid i[U(\cdot) \psi]^{\prime}(0) \text { exists}\right\}.
\end{equation}

\textbf{Stone's theorem} says that if we start from a UAOPG, which could correspond to translation, rotation etc, we can obtain self-adjoint generators corresponding to quantum mechanical observables , momentum, angular momentum etc. In other words, for a UAOPG $U(\cdot)$, its generator $A: \mathcal{D}^S_A \rightarrow \mathcal{H}$ is self-adjoint on $\mathcal{D}_A^S$. The generators of the group may not be defined on the whole of the Hilbert space, but on the Stone domain they are self-adjoint (observable). Moreover, we can reconstruct the group from its generators as:
\begin{equation}
    U(t) = e^{-itA}
\end{equation}

Proving Stone's theorem \cite{Hall} enables us to verify the following:
\begin{enumerate}[i]
    \item  Generator A is densely defined, i.e the domain of A lies in the Hilbert space. From that we can see that A* is also well defined.
    \item A is symmetric and essentially self-adjoint.
    \item $U(t) = e^{-itA^{**}}$, from which we can conclude that A is the closure of itself, A=A**.
\end{enumerate}
\begin{example}\textbf{Position operator.}
For a Hilbert space that's defined to be $\mathcal{H}=L^2(\mathbb{R}^3)$, then the group $U(t): \mathcal{H} \rightarrow \mathcal{H}$ would have generator $\hat x$. The action of $U(t)$ on vector $\psi$ would introduce a phase:
\begin{equation}
    (U(t) \psi) (x) := e^{-itx} \psi (x)
\end{equation}
and we can see that (1) $U(t)U(s) = U(t+s)$, (2) $U(0) = id_{\mathcal{H}}$, and (3) $|| U(t) \psi|| = ||\psi||$. Thus, by Stone's theorem, the generator: $\hat x : \mathcal{D}^S_{\hat x} \rightarrow \mathcal{H}$  is self-adjoint on the domain $\mathcal{D}^S_{\hat x}$. It acts on a vector $\psi\in\mathcal{D}^S_{\hat x}$ as:
\begin{equation}
    (\hat x \psi) (x) := x \psi (x)
\end{equation}
and we clearly see that we have constructed the position operator.
\end{example}

\subsection{Stone-von Neumann Theorem} \label{s vn}

The Stone-von Neumann uniqueness theorem, generalizes Stone's theorem to a \textit{pair} of self-adjoint operators \cite{v1931eindeutigkeit}.
It states that there is a unique way to represent the canonical commutation relations on a finite-dimensiona Hilbert space, up to unitary equivalence. Therefore, there are no other interesting representations of this algebra.  theorem only applies to systems with a finite number of degrees of freedom This result is essential for ensuring the consistency of quantum mechanics. Proof can be found in \cite{Hall}.

For example, in position representation, $\hat{x}$ is represented as a multiplication operator, as shown in (\ref{prep1}), and $\hat{p}$ is represented as the differential operator, as shown in (\ref{prep2}). The canonical commutation relation would be:
\begin{equation} \label{comm}
    [\hat{x} , \hat{p} ] = [x \, , -i \hbar \, \frac{d}{dx}] = i \hbar,
\end{equation}
whereas if we wish to switch the roles of $\hat{x}$ and $\hat{p}$, making the the former the the differential operator, as in (\ref{mrep1}) and the later the multiplication operator, as in (\ref{mrep2}), we essentially go to the momentum representation, the canonical commutation would be:
\begin{equation}
    [\hat{x} , \hat{p} ] = [i \hbar \, \frac{d}{dp} \, , \,  p] = i \hbar
\end{equation}

\begin{definition}
    An irreducible representation is one that cannot be further decomposed into smaller representations that are still irreducible.
\end{definition}
For example, consider the group of rotations in three-dimensional space, denoted by SO(3). A representation of SO(3) associates a matrix with each element of the group, in such a way that the group structure is preserved. An irreducible representation of SO(3) is one that cannot be written as a direct sum of smaller representations, and in this case, it corresponds to a matrix that cannot be block-diagonalized into smaller matrices.\\

We note here that Stone-von Neumann theorem establishes the equivalence of irreducible representations of the canonical commutation relations in QM. However, this theorem does not extend to QFT, where there are multiple inequivalent irreducible representations. This leads to a much richer algebraic structure in QFT than in QM.

\subsection{Fourier link between $\phi(x)$ and $\psi(p)$ spaces} \label{Fourier link}

Fourier transformation provides a map between two parallel descriptions of quantum mechanics. It provides a way to construct different representations of the CCR, which can be shown to be unitarily equivalent to each other. Abstractly, the quantum state is represented by a \textit{state vector} $| \psi \rangle $. The momentum wavefunction $\psi(p)$ and position wavefunction $\phi(x)$ are two different representations of that same vector in the momentum and position bases, respectively. The Fourier transform is a unitary operator that maps the position and momentum operators in one representation to a new set of operators in a different representation.
\begin{equation}
    \phi(x) = \frac{1}{\sqrt{2 \pi}} \int_{-\infty}^{\infty} dp \, e^{ipx} \psi(p)
\end{equation}
and the inverse Fourier transformation of the wavefunction:
\begin{equation}
     \psi(p)= \frac{1}{\sqrt{2 \pi}} \int_{-\infty}^{\infty} dx \, e^{-ipx} \phi(x)
\end{equation}
This clearly illustrates that there's a 1-1 correspondence between the $|x \rangle$ and $|p \rangle$.

By applying the Fourier transform repeatedly, we can construct an infinite number of representations of the CCR, each related to the previous one by a unitary transformation. However, as the Stone-von Neumann theorem states, all irreducible representations of the CCR on a finite-dimensional Hilbert space are unitarily equivalent. Therefore, in the case of finite-dimensional Hilbert spaces, the Fourier transform cannot give rise to inequivalent representations of the CCR. This is because all irreducible representations of the CCR are equivalent to the Schrödinger representation\cite{Gieres_2000}, which is uniquely defined.

In the case of infinite-dimensional Hilbert spaces, however, the situation is more complicated, and the Stone-von Neumann theorem does not apply. In this case, the Fourier transform can give rise to inequivalent representations of the CCR, and the choice of representation can have physical consequences.

\chapter{Algebraic Quantum Formalism}\label{3}
In this section, we will  examine the ``algebraic approach" to quantum mechanics, which is a method that delves into the principles of quantum mechanics through the lens of the mathematical theory of algebras of operators. The concept of Algebraic Quantum Mechanics was developed to provide a mathematically rigorous theory for quantum systems with an infinite number of degrees of freedom, such as those encountered in quantum field theory and  quantum statistical mechanics. This approach goes beyond the traditional Hilbert space framework and utilizes operator algebras, such as von Neumann algebras and C*-algebras, to describe the observables of the system.

The algebraic approach to quantum mechanics places an emphasis on understanding the abstract structure of the set of observables, which are the measurable quantities of a system, as well as the set of states, which describe the possible configurations of a system. In this framework, we are particularly interested in the rules governing the probabilities associated with the measurements of observables over states, which dictate how the system behaves when an observable is measured.

By employing the algebraic approach, we can gain deeper insights into the underlying principles of quantum mechanics and develop a more solid foundation for understanding the often counter-intuitive behavior of quantum systems. This approach also enables us to analyze quantum mechanics from a purely mathematical standpoint, which can prove invaluable when attempting to solve complex problems or develop new theories in the field.\\

\section{Operator Algebras}

In quantum mechanics, the observables of a system are represented by operators acting on a state $| \psi \rangle$ in the Hilbert space $\mathcal{H}$. In QFT, since the system has an infinite number of degrees of freedom, there are an infinite number of observables. The operator algebra approach provides a mathematical framework for studying these observables. It is possible to describe to quantum systems without directly referring to a vector space. Quantum mechanics has no standard basis, so working with the abstract operators directly can give more insight than representing them in a Hilbert space. The observables of a system can be organized into an algebra, called the algebra of observables, which includes the identity operator, linear combinations of observables, and products of observables.

The operator algebra is important because it provides a rigorous mathematical structure for studying QFT.  It also allows us to study the algebraic properties of observables, such as commutation and anti-commutation relations, which are important for understanding the behavior of the system. Additionally, the operator algebra approach provides a powerful tool for studying the properties of QFT on curved spacetimes and in the presence of external fields.

The algebraic structure of QFT is closely related to von Neumann algebras, which are a type of operator algebra that arise naturally in the study of QFT. These algebras play a central role in the operator algebra approach to QFT, as they provide a framework for studying the algebraic properties of observables. The algebra of observables refers to the mathematical structure which describes the set of observables of a quantum system and the relations between them. In this we go into characterizing this algebra into different types of von Neumann algebras, their properties will be detailed.

\section{Commutants}

For an algebra $\mathcal{A} \subseteq \mathcal{B(H)}$, its commutant $\mathcal{A'}$ is defined as the set of all bounded operators that commute with with $\mathcal{A}$
\begin{equation}
    \mathcal{A'} = \{B \in \mathcal{B(H)} | [A,B] = 0 , \forall A \in \mathcal{A}\}
\end{equation}
\begin{remark}
   For a a bounded algebra $\mathcal{B(H)}$ with dimensions $d$, there exists a trivial commutant that's isomorphic to the complex numbers.
\end{remark}
\begin{example} Consider $\mathcal{H} = \mathbb{C}^3$, then the bounded operators $\mathcal{B(H)} = \mathcal{M}_3$. The most general for of $\mathcal{M}_3$ is: 
\begin{equation}
    \mathcal{M}_3 =
    \begin{pmatrix}
    a & b & c\\
    d & e & f\\
    g & h & i
    \end{pmatrix}
\end{equation}
Then  its commutant $\mathcal{M}'_3$ is:
\begin{equation}
    \mathcal{M}'_3 =
    \begin{pmatrix}
    u & 0 & 0\\
    0 & u & 0\\
    0 & 0 & u
    \end{pmatrix}
    = u \begin{pmatrix}
    1 & 0 & 0\\
    0 & 1 & 0\\
    0 & 0 & 1
    \end{pmatrix}
    = \mathbb{C} \; \mathbb{I}_3
\end{equation}
Where the elements of the matrix $\in \mathbb{C}$.\\
\end{example}
\begin{remark}
    If $\mathcal{A}$ is commutative, then $\mathcal{A} \subseteq \mathcal{A}'$ and $\mathcal{A} \subseteq \mathcal{A}''$
\end{remark}
\begin{example}
Consider an algebra $\mathcal{A} \subset \mathcal{M}_3$, where $\mathcal{A}$ is commutative. Then for
\begin{equation} \label{a commutant}
    \begin{pmatrix}
    a & 0 & 0\\
    0 & a & 0\\
    0 & 0 & b
    \end{pmatrix}
    \in \mathcal{A}
\end{equation}
Then its commutant $\mathcal{A}'$ has the form
\begin{equation}
    \mathcal{A}' =
    \begin{pmatrix}
    a & b & 0\\
    c & d & 0\\
    0 & 0 & v
    \end{pmatrix}  \cong \mathcal{M}_2 \otimes \mathbb{C}
\end{equation}
We clearly see that that $\mathcal{A} \subseteq \mathcal{A}'$.\\
\end{example}
\begin{remark}
    If $\mathcal{A} = \mathbb{C \; I}$, then $\mathcal{A}' = \mathcal{B(H)}$.
\end{remark}
\begin{example}Consider a general element in $\mathcal{A}$ such that:
\begin{equation} \label{unitary commutant}
     \mathcal{A} \ni \begin{pmatrix}
    a & 0 & 0\\
    0 & a & 0\\
    0 & 0 & a
    \end{pmatrix}
    \cong \mathbb{C} 
\end{equation}
Then its commutant is:
\begin{equation} 
     \mathcal{A}' \ni \begin{pmatrix}
    a & b & c\\
    d & e & f\\
    g & h & i
    \end{pmatrix}
    \cong \mathcal{M}_3
\end{equation}
We see that $\mathcal{A}' = \mathcal{B(H)}$. \\
\end{example}

Providing a preview of the forthcoming content in section (\ref{vn}): if we repeat the above process, we see that  $\mathcal{A} \subseteq \mathcal{A}''$. If the equality holds we call the algebra satisfying \textit{bicommutivity} $\mathcal{A} = \mathcal{A}''$ von Neumann algebra.

\section{Von Neumann Algebra} \label{vn}
\begin{remark}
A subalgebra $\mathcal{A}\subseteq\mathcal{B}(\mathcal{H})$ is a von Neumann algebra iff it's bicommutatant $\mathcal{A}''=\mathcal{A}$.
\end{remark}
Von Neumann algebra play a central role in axiomatic approaches to quantum field theory and statistical mechanics. The underlying concept is that the observables in a physical theory should exhibit an algebraic structure. In the context of QFT, von Neumann algebras play a crucial role in describing the algebraic structure of observables \cite{sorce2023notes}.

The \textit{von Neumann algebra} is C*-subalgebra of the bounded linear operators on a Hilbert space, closed in the weak operator topology. Given a subset $\mathcal{A}$ of the bounded operators $\mathcal{B(H)}$, the \textit{von Neumann algebras} can be defined as the set that satisfy the following: 
\begin{enumerate}[i]
    \item $\mathcal{A}$ is an algebra: $\mathcal{A}$ is closed under addition, operator and scalar multiplication, and contains the identity.
    \item $\mathcal{A}$ is a *-algebra: $\mathcal{A}$ is closed under hermitian conjugation.
    \item $\mathcal{A}$ is closed with respect to the strong operator topology: there exists a sequence of operators $a_n$ that converges to an operator $a$ as $n \rightarrow \infty$ 
    \begin{equation}
    \exists a_n \in A \quad \text { s.t. } \forall|\psi\rangle \quad \lim _{n \rightarrow \infty} a_n|\psi\rangle=a|\psi\rangle \quad \Rightarrow \quad a \in A
\end{equation}
\end{enumerate}

Additionally, if the \textit{center} of $\mathcal{A}$ is trivial, i.e. it consists of only of complex numbers, we say that $\mathcal{A}$ is a \textit{von Neumann factor}. The center of a von Neumann algebra is the set of all elements that commute with every element in the algebra. A von Neumann algebra is called a ``factor" if its center consists only of scalar multiples of the identity operator. The nice thing about factors is that they are the building-blocks of von Neumann algebra, as any von Neumann algebra that are not factora can be written as a direct sum/integral of factors. There von Neumann factors can be classified to three types (Type I, Type II, and Type III) based on their properties and representation theory \cite{CYRIL}. A rough classification is summarized in the table below:\\

\begin{tabular}{ |p{1.8cm}||p{4cm}|p{4cm}|p{3.6cm}|  }
 \hline
 \rowcolor{lightgray}
 \multicolumn{4}{|c|}{von Neumann factors} \\
 \hline
 
 Types & Projectors & Trace & Irreducible representation in  $\mathcal{H}$\\
 \hline
 
Type $I_d$ & Minimal projectors & Defined & Irreducible\\
Type $I_\infty$ & Minimal projectors & Undefined & Irreducible\\
 \hline
 Type $II_1$ &  Fintite \& non-minimal & Defined but not unique & Does not exist\\
  Type $II_\infty$ &  Fintite \& non-minimal  & Defined but not unique & Does not exist\\
  \hline
 Type III   & Non-minimal & Undefined & Does not exist\\
 \hline
\end{tabular}\\

Factors are particularly interesting because they have unique properties that are not shared by non-factor von Neumann algebras. The classification of von Neumann algebras is important in QFT because it is related to the structure of spacetime.

\subsection{Type I}

Type $I$ von Neumann algebras are the simplest type and include finite and infinite factors. They arise in the algebraic description of quantum mechanics and are closely related to the Stone-von Neumann theorem. 

\textit{Type $I$ factors} are the algebra of all bounded operators that act irreducibly on $\mathcal{H}$, either on $\mathcal{H}$ itself, or some smaller subsystem $\mathcal{H_A}$ such that $\mathcal{H} = \mathcal{H_A} \otimes \mathcal{H_B}$. They can be labelled by the dimension $d$ of the Hilbert space  $\mathcal{H}$. There is a unique (up to isomorphism) Type $I_d$ factors for $d \in \mathbb{N}$. In the case that $\mathcal{H}$ has infinite dimensions, the algebra of bounded operators is of Type $I_\infty$. Moreover, Type $I$ factors have \textit{minimal projector}, i.e. generated by a single projection (an operator that is its own square and is self-adjoint) and are therefore commutative. The uniqueness of the projection property of type $I$ factors is due to the fact that they have a unique cyclic and separating vector, which is not the case for Type $II$ and Type $III$ factors. The cyclic and separating vector property of type $I$ factors allows for the construction of a projection valued measure, which is used to define the density matrix. This property is not present in Type $II$ and Type $III$ factors, which means that the density matrix cannot be defined in the same way. Instead, in type $II$ and Type $III$ factors, one needs to use the Tomita-Takesaki  theory and the associated modular theory to define the density matrix. These concepts will be discussed in detail in the following sections. 

In finite dimensions, as is the case of quantum mechanics, the only possible von Neumann factors are Type $I$ factors. They are used to describe systems such as spin chains and lattice models. However, as we move to QFT, the more complex Type $II$ and Type $III$ von Neumann algebras become relevant.

\subsection{Type II}

Type II factors have no minimal projectors, but finite\footnote{A finite von Neumann projector is one that has finite rank, meaning that it projects onto a subspace of finite dimension. In contrast, an infinite von Neumann projector would project onto a subspace of infinite dimension.} projectors.
They can be classified to  \textit{hperfinite Type $II_1$ factors} and \textit{Type $II_\infty$ factors}. In physics, infinite structures are often obtained as limits of finite structures, and therefore, the algebras that commonly emerge in physics are predominantly hyperfinite\footnote{A von Neumann algebra is said to be hyperfinite if it can be approximated arbitrarily well by finite-dimensional von Neumann algebras.} ones. The concept of hyperfinite algebras is important because it provides a rigorous mathematical framework for dealing with infinite-dimensional systems. For example, in quantum field theory, we often work with infinite-dimensional Hilbert spaces, but these can be difficult to handle mathematically. By using hyperfinite algebras to approximate these spaces, we can apply techniques from finite-dimensional linear algebra to study them.

Type $II_\infty$ can be constructed by the tensor product: Type $II_\infty = II_1 \otimes I_\infty$. We can construct hyperfinite Type $II_1$ by considering a vector space ${V}$ of $2 \times 2$ complex matrices which can be given a Hilbert space structure as we shall see below. %that consists of infinitely many pairs of entangled qubits \cite{witten2018}.
\begin{equation}
    v = \begin{pmatrix}
    1 & 0 \\
    0 & 1 \\
    \end{pmatrix} = 
    \begin{pmatrix}
    1 \\
    0 \\
    \end{pmatrix}\begin{pmatrix}
    1 & 0 \\
    \end{pmatrix}  +
    \begin{pmatrix}
    0 \\
    1 \\
    \end{pmatrix}\begin{pmatrix}
    0 & 1 \\
    \end{pmatrix} =
    \begin{pmatrix}
    1 \\
    0 \\
    \end{pmatrix} \otimes \begin{pmatrix}
    1 & 0 \\
    \end{pmatrix}  +
    \begin{pmatrix}
    0 \\
    1 \\
    \end{pmatrix} \otimes \begin{pmatrix}
    0 & 1 \\
    \end{pmatrix} = v_1 + v_2  \; \in V
\end{equation}
where $v,v_1,v_2\in V$ and they represent 2-qubit compound states; $v_1$ and $v_2$ are separable pure states and $v$ is an un-normalized entangled pure state.   To turn the vector space space $V$ into a Hilbert space structure, we can define an inner product:
\begin{equation} \label{trace}
    \langle v_1, v_2\rangle=\operatorname{Tr} [v_1^{\dag} v_2]
\end{equation}
We can define matrices $a \in M_2$ that act on  $V$ from the left,
and $b \in M'_2$ that act on $V$ from the right as such:
    \begin{align}
        & a (V) = a V\\
        & b (V) = V b^\dag
    \end{align}
For a maximally entangled vector $v \in V$, we define the normalized state $I'_2 = \frac{1}{\sqrt{2}} I_2$.

Now we are ready to construct an infinite tensor product pre-Hilbert space $\mathcal{H}_0$\footnote{A pre-Hilbert space is a vector space equipped with an inner product that is not necessarily complete. The completion of a pre-Hilbert space with respect to its norm yields a Hilbert space. } as follows. %space of these qubits using the tensor product:
Consider the infinite product
\begin{equation}
         v_1 \otimes v_2 \otimes \cdots \otimes v_k \otimes \cdots \in V^{[1]} \otimes V^{[2]} \otimes \cdots \otimes V^{[k]} \otimes \cdots
\end{equation}
where  finitely many $v_k$'s are NOT equal to $I'_2$ and the infinite remaining $v_k$'s  are $I'_2$. If $v_k\neq I'_2$ then it is an arbitrary $2 \times 2$ matrix. For example, elements $v\in\mathcal{H}_0$ can be like: 
\begin{comment}
        \begin{align}
            v = & \quad \frac{1}{\sqrt{2}} I &&\otimes  \quad \frac{1}{\sqrt{2}} I  \quad \quad \quad \otimes \cdots \otimes \quad v_k  \quad \otimes \cdots \in V^{[1]} \otimes V^{[2]} \otimes \cdots \otimes V^{[k]} \otimes \cdots  \nonumber \\
            = & \frac{1}{\sqrt{2}} \begin{pmatrix}
            1 & 0 \\
             0 & 1 \\
            \end{pmatrix} && \otimes \frac{1}{\sqrt{2}} \begin{pmatrix}
            1 & 0 \\
            0 & 1 \\
            \end{pmatrix} \otimes \cdots \otimes \begin{pmatrix}
            a & b \\
            c & d \\
            \end{pmatrix}  \otimes \cdots \in V^{[1]} \otimes V^{[2]} \otimes \cdots \otimes V^{[k]} \otimes \cdots
        \end{align}
\end{comment}
\begin{equation}
     v =  I'_2\otimes I'_2\otimes \begin{pmatrix}
            a_1 & b_1 \\
            c_1 & d_1 \\
            \end{pmatrix} \otimes I'_2 \otimes\begin{pmatrix}
            a_2 & b_2 \\
            c_2 & d_2 \\
            \end{pmatrix} \otimes I'_2  \cdots 
\end{equation}
Note that this is not quite a Hilbert space, but a countably infinite pre-Hilbert space. We need to \textit{complete} it in order to obtain $\mathcal{H}\supset\mathcal{H}_0$ To do that, we consider $v=v_1 \otimes v_2 \otimes \cdots$  and  $w=w_1 \otimes w_2 \otimes \cdots$  in $\mathcal{H}_0$. We know that there are only finite amount $n$ of vectors $v_k, w_k$ that are not $I'_2$, so we find a find $k > n$ and truncate at $n$ so that every non-identity element is included; $v_{\langle n\rangle}=v_1 \otimes v_2 \otimes \cdots \otimes v_n$, $w_{\langle n\rangle}=w_1 \otimes w_2 \otimes \cdots \otimes w_n$. Note that these are finite dimensional. Similar to the trace definition eq (\ref{trace}):
\begin{equation}
    \langle v, w\rangle: = \operatorname{Tr} [v_1^\dag w_1 ] \; \operatorname{Tr} [v_2^\dag w_2]  \cdots  \operatorname{Tr} [v_n^\dag w_n] \operatorname{Tr}[ I'^\dag_2 I'_2] \cdots=\operatorname{Tr} [v_{\langle n\rangle}^{\dagger} w_{\langle n\rangle}]<\infty
\end{equation}
Where $\operatorname{Tr} [I'^\dag_2 I'_2 ]= 1$. Finally, for $v = v_1 \otimes v_2 \otimes \cdots \otimes v_n \otimes \cdots$ to be restricted in tensor product, $v_n$ must tend to $I'_2$ as $n \rightarrow \infty$.

Now we follow a similar procedure for the algebra $\mathcal{A}$. The goal is to define the algebra as an infinite tensor product $M_2^{[1]} \otimes M_2^{[2]} \otimes \cdots \otimes M_2^{[n]} \otimes \cdots$. We note that a general element $a \in \mathcal{A}$ does not have finitely many $a_k \neq I_2$. For this reason, we have the problem that the action of the algebra takes us out of the Hilbert space, $a \mathcal{H} \nsubseteq \mathcal{H}$. To solve this problem, we define $\mathcal{A}_0$ such that it contains finitely many $a_k \neq I_2$. Now the action of this algebra is constrained within the Hilbert space, $\mathcal{A}_0 \mathcal{H} \subseteq \mathcal{H}$. Finally we add the limit to close  $\mathcal{A}_0$ to get $\mathcal{A}: \mathcal{H} \rightarrow \mathcal{H}$.
The commutant of $\mathcal{A}$ is $\mathcal{A}'$, and it includes elements $b \in M_{2}^{'[1]} \otimes M_{2}^{'[2]} \otimes \cdots \otimes M_{2}^{'[n]} \otimes \cdots$ that act on $v$ from the right.

We introduce natural linear function $F(\mathrm{a})=\langle\Psi|\mathrm{a}| \Psi\rangle$ on the algebra $\mathcal{A}$. This will prove to be an important tool for dealing with entanglement in quantum field theory, as  the function $F(\mathrm{a})$ allows us to calculate expectation values of observables and study their entanglement properties. We define a vector:
\begin{equation}
    \Psi=I_2^{\prime} \otimes I_2^{\prime} \otimes \cdots \otimes I_2^{\prime} \otimes \cdots \in \mathcal{H}
\end{equation}
Then the linearized function has the properties of a trace, $F(\mathrm{ab})= F(\mathrm{ba})$. Plugging ($\mathrm{a} \; \mathrm{b}$) yields:
\begin{equation}
    F(\mathrm{ab})=\operatorname{Tr}_{M_2^{[1]} \otimes  \cdots \otimes M_2^{[k]}} \mathrm{a}_1 \mathrm{b}_1 \otimes \cdots \otimes \mathrm{a}_k \mathrm{b}_k= \operatorname{Tr}_{M_2^{[1]} \otimes  \cdots \otimes M_2^{[k]}}\mathrm{b}_1 \mathrm{a}_1  \otimes \cdots \otimes \mathrm{b}_k \mathrm{a}_k =F(\mathrm{ba})
\end{equation}
Where the trace now is normalized so that $\operatorname{Tr} 1 = 1$.\\
\begin{remark}
    Type II factors have no unique trace, because we can multiply trace  by a constant. But the difference between entropy can be defined, because the constants will cancel. So simply a notion of trace is sufficient, as we will see in section \ref{ent entropy}.
\end{remark}

\subsection{Type III}

Type III algebra is the most general class of factors. It has no trace and no minimal projectors. We can start this generalization by switching the maximally entangled pair of qubits $I'_2$ in Type $II$ factors with nonmaximally entangled quibit pair $k_{2,\lambda}$, such that:
\begin{equation}
    k_{2, \lambda}=\frac{1}{(1+\lambda)^{1 / 2}}\left(\begin{array}{cc}
1 & 0 \\
0 & \lambda^{1 / 2}
\end{array}\right) \in V
\end{equation}
where $0 < \lambda < 1$. 
After replacing the normalized identities with this new matrix, the pre-Hilbert  space $\mathcal{H}_0$ is spanned by vectors $v_1 \otimes v_2 \otimes \cdots \otimes v_n \otimes k_{2, \lambda}\otimes k_{2, \lambda} \cdots \in V^{[1]} \otimes V^{[2]} \otimes \cdots$. Because this space has a structure like that of a Hilbert space, we can use our definition of the dot product and trace in eq(\ref{trace}):
\begin{equation}
    \langle k_{2, \lambda}, k_{2, \lambda}\rangle=\operatorname{Tr} [k_{2, \lambda}^{\dag} k_{2, \lambda}] = \operatorname{Tr} \left[\frac{1}{1+\lambda}\left(\begin{array}{cc}
1 & 0 \\
0 & \lambda
\end{array}\right)\right] =1
\end{equation}
The next step would be to close $\mathcal{H}_0$ with respect to the inner product to obtain the complete Hilbert space $\mathcal{H_\lambda}$. We do this by truncating, exactly as we did in Type $II$.
\begin{equation}
    \langle v, w\rangle: = \operatorname{Tr} [v_1^\dag w_1 ] \; \operatorname{Tr} [v_2^\dag w_2]  \cdots  \operatorname{Tr} [v_n^\dag w_n] \operatorname{Tr}[ k_{2, \lambda}^\dag  k_{2, \lambda}] \cdots=\operatorname{Tr} [v_{\langle n\rangle}^{\dagger} w_{\langle n\rangle}]<\infty
\end{equation}
Similarly, we take the same algebra $\mathcal{A}_0$ that we started with in Type $II$, then complete it with respect to $\mathcal{H_\lambda}$ to obtain $\mathcal{A}_\lambda$ and $\mathcal{A}'_\lambda$.

Finally, similar to what was done in Type $II$, we define natural linear function $F(\mathrm{a})=\langle\Psi_{\vec{\lambda}}|\mathrm{a}| \Psi_{\vec{\lambda}}\rangle$ on the algebra $\mathcal{A}_\lambda$. 
\begin{remark}
    The difference between between Type $II$ and Type $III$ algebra is that the later does NOT admit a trace. In essence, $F(\mathrm{ab}) \neq F(\mathrm{ba})$.
\end{remark}
Remember, previously we had $F(\mathrm{ab}) = \langle\Psi|\mathrm{ab}| \Psi\rangle = \operatorname{Tr} [I'_2 \mathrm{a}_{\langle k\rangle} \mathrm{b}_{\langle n\rangle} I'_2] = F(\mathrm{ba})$. This was allowed because $\Psi \in \mathcal{H}$ was a tensor product of infinite normalized identity matrices. This time we are dealing with a different space, namely $\mathcal{H_\lambda}$, we redefine $\Psi$ as:
\begin{equation}
    \Psi_{\vec{\lambda}}=k_{2, \lambda_1} \otimes k_{2, \lambda_2} \otimes \cdots \otimes k_{2, \lambda_n} \otimes \cdots \in \mathcal{H_\lambda}
\end{equation}
Now the trace is not well-defined. 
\begin{equation}
    F(\mathrm{ab}) = \langle\Psi_{\vec{\lambda}}|\mathrm{ab}| \Psi_{\vec{\lambda}}\rangle = \operatorname{Tr} [k_{2, \lambda} \mathrm{a}_{\langle k\rangle} \mathrm{b}_{\langle n\rangle} k_{2, \lambda}] \neq  \operatorname{Tr} [k_{2, \lambda} \mathrm{b}_{\langle k\rangle} \mathrm{a}_{\langle n\rangle} k_{2, \lambda}] = F(\mathrm{ba})
\end{equation}

Type $III$ factors can be further classified based on how the sequence of $\lambda$ converge\cite{araki1968classification}:
\begin{itemize}
    \item 
    Type III: $k_{ \lambda_1} \otimes k_{\lambda_2} \otimes \cdots \otimes k_{\lambda_n}$, where $0<\lambda_i<1$.
    \item Type $III_0$: $\lambda_i \rightarrow 0$. If the sequence of $\lambda_i$ converges to zero fast enough, we get Type $I_\infty.$
    \item Type $III_\lambda$: $\lambda_i \rightarrow \lambda$. For example, $\lambda_i = \lambda \quad \forall i$, where we have finitely many $\lambda_i \neq \lambda$. A special case is when $\lambda_i \rightarrow 1$, we obtain Type $II$ factors.
    \item  Type $III_1$: here $\lambda_i$ does not converge. A simple case is $\lambda_i \in\{\lambda, \tilde{\lambda}\}$, where we have infinitely many of each. The span of this type is $[0, \infty)$. 
\end{itemize}

\section{Tomita-Takesaki Theory}

The Tomita-Takesaki theorem is a fundamental result in the theory of von Neumann algebras, which provides a deep connection between the algebraic and geometric structures of these mathematical objects \cite{Takesaki:1970aki}. In particular, the theorem establishes a canonical way of constructing a modular automorphism group associated with any von Neumann algebra acting on a Hilbert space.

More precisely, given a von Neumann algebra $\mathcal{A}$ acting on a Hilbert space $\mathcal{H}$, the Tomita-Takesaki theorem asserts the existence of a modular conjugation operator $J$ and a modular operator $\Delta$ satisfying certain axioms. These operators are intimately related to the geometry of the underlying Hilbert space and capture important physical features such as time evolution and symmetry transformations.

In addition to its intrinsic mathematical interest, the Tomita-Takesaki theorem has numerous applications in quantum field theory, statistical mechanics, and other areas of physics. For example, it plays a crucial role in the analysis of thermal equilibrium states and the formulation of the Kubo–Martin–Schwinger (KMS) condition, which characterizes the behavior of quantum systems at finite temperatures \cite{Takesaki:1973ge}. The theorem also provides a powerful tool for investigating the structure of quantum entanglement and the emergence of classical physics from quantum mechanics \cite{Strocchi2008AnIT, chandrasekaran2023large}. Moving forward, we rely Witten's construction of the theorem in \cite{witten2018}.

\subsection{Tomita modular operator} \label{tomita}

Another important concept that we will  use to define entropy in section (\ref{ent entropy}) is the Tomita operator. We will classify the algebra that define QFT, the algebra of QFT don’t admit a trace, so we need a new notion to incorporate undefined matrices in a defintion of entropy. That’s one of the uses of the Tomita operator and the modular operator. The Tomita operator can be defined as:
\begin{equation}
    S_{\Psi} a|\Psi\rangle=a^{\dagger}|\Psi\rangle
\end{equation}
Where $S_{\Psi}$ is an antilinear operator acting on a state $\Psi$ that's cyclic and separating for the algebra, and $a \in \mathcal{A}$. 
\begin{itemize}
    \item \textit{cyclic}: $\mathcal{A}\Psi$ is dense\footnote{Dense means that any vector $\Phi\in\mathcal{H}$ can be approached arbitrarily closely (with respect to the norm induced by the inner product) by a sequence of vectors $\Phi_n$ in $\mathcal{A}\Psi$. \cite{Melrose}} in $\mathcal{H}$. 
    \item\emph{separating}: $a\Psi=0\Leftrightarrow a=0$ \cite{Melrose}.
\end{itemize}
If we investigate the properties of $S$, we see that $S_{\Psi}$ is invertible:
\begin{equation}
    S_{\Psi}^2 = 1
\end{equation}
Hence,
\begin{equation}
    S_{\Psi}|\Psi\rangle=|\Psi\rangle
\end{equation}.
Recall that if $W$ is a \textit{linear} operator, then its action on states $\Psi, \chi$:
\begin{equation}
    \langle\Psi \mid W ^\dag\chi\rangle=\left\langle W\psi | \chi\right\rangle
\end{equation}
and if $W$ is \textit{antilinear}, then it acts as:
\begin{equation}
     \langle\Psi \mid W ^\dag\chi\rangle=\overline{\left\langle W \Psi \mid \chi\right\rangle}=\langle\chi|W\psi\rangle
\end{equation}
Since $S$ is antilinear, we will be making use of this property quite often.
Furthermore, for $S_{\Psi}$ of an algebra $\mathcal{A}$, we can define $S_{\Psi}'$ for the commuting algebra $a'\in\mathcal{A}'$. As it turns out \cite{witten2018}:
\begin{equation} \label{s conj}
    S_{\Psi}^{\dagger} = S_{\Psi}^{\prime} 
\end{equation}

The Tomita operator has plenty of interesting properties. Knowing that it is invertible, we can further investigate it by polar decomposition it to an antiunitary operator $J_{\Psi}$ and a unitary, positive-definite hermitian operator $\Delta_{\Psi}=S_{\Psi}^\dag S_{\Psi}$:
\begin{equation} \label{polar dec}
    S_{\Psi}=J_{\Psi} \Delta_{\Psi}^{1 / 2} \quad , \quad  S_{\Psi}^{\dagger} = \Delta_{\Psi}^{1 / 2}  J_{\Psi}
\end{equation}
Where $J_{\Psi}:\mathcal{A}\rightarrow\mathcal{A}'$ is the \textit{modular conjugation}, and $\Delta_{\Psi}: \mathcal{A}\rightarrow\mathcal{A}$ is the \textit{modular operator}. The unitary part maps an element of the algebra onto itself, wheras the antiunitary part maps an element of the algebra onto its commutant. This implies that:
\begin{equation}
    S_{\Psi}^{\dagger} S_{\Psi} = \Delta_{\Psi}^{1 / 2} J_{\Psi}^2\Delta_{\Psi}^{1 / 2} = \Delta_{\Psi} 
\end{equation}
and
\begin{equation}
    ( S_{\Psi} S_{\Psi}^{\dagger})(S_{\Psi}^{\dagger} S_{\Psi}) = S_{\Psi} (S_{\Psi}^{\dagger} S_{\Psi}^{\dagger} ) S_{\Psi} = S_{\Psi} (1) S_{\Psi} = S_{\Psi}^2 =1 
\end{equation}
which implies $(S_{\Psi} S_{\Psi}^{\dagger}) = (S_{\Psi}^{\dagger} S_{\Psi})^{-1}$, i.e  $(S_{\Psi} S_{\Psi}^{\dagger}) = (\Delta_{\Psi})^{-1}$. It follows that:
\begin{equation}
    S_{\Psi} S_{\Psi}^{\dagger} = J_{\Psi} \Delta_{\Psi}^{1 / 2} \Delta_{\Psi}^{1 / 2} J_{\Psi} =  J_{\Psi} \Delta_{\Psi} J_{\Psi} = \Delta_{\Psi}^{-1}
\end{equation}
In addition to these properties, we can see that $S_{\Psi}^2 = (J_{\Psi} \Delta_{\Psi}^{1 / 2}) (J_{\Psi} \Delta_{\Psi}^{1 / 2}) = 1$, 
it's straightforward to deduce  $(J_{\Psi} \Delta_{\Psi}^{1 / 2} J_{\Psi} ) \Delta_{\Psi}^{1 / 2} = 1 \rightarrow J_{\Psi} \Delta_{\Psi}^{1 / 2} J_{\Psi}  = \Delta_{\Psi}^{-1 / 2}$.

From the definition of $S$, $S_{\Psi} \Psi = S_{\Psi}^\dag \Psi = \Psi$. From here we can see that:
\begin{equation}
    \Delta_{\Psi} |\Psi\rangle = S_{\Psi}^{\dagger} S_{\Psi} |\Psi\rangle = S_{\Psi}^{\dagger} |\Psi\rangle = |\Psi\rangle
\end{equation}
So the eigenvalue of $\Delta_{\Psi}$ is simply 1. We can generalize this for any function $f$:
\begin{equation}
    f(\Delta_{\Psi}) |\Psi\rangle = f(1) |\Psi\rangle
\end{equation}
For example, if $f(x) = e^{xt}$, then $e^{\Delta_{\Psi}t} |\Psi\rangle= e^{t} |\Psi\rangle$.

In quantum mechanics, the dynamics are governed by the evolution operator $U(t) = e^{iHt}$. Loosely speaking, we ``generalize" this to $\Delta^{it}$. In essence, replacing $e^{H}$ by $\Delta$ \cite{Vaughan}.

\subsection{Relative modular operator}

So far we have looked at the Tomita operator $S$ acting on the state $\Psi$. A generalization of $S_{\Psi}$ can relate $\Psi$ to another `state' $\Phi$. We define the \textit{relative Tomita operator} $S_{\Psi \mid \Phi}$ for the algebra $a \in \mathcal{A}$:
\begin{equation}
    S_{\Psi \mid \Phi} a|\Psi\rangle= a^{\dagger}|\Phi\rangle
\end{equation}
We defined $\Psi$ to be cyclic separating for the algebra. The state $\Phi$ is arbitrary. If it is cyclic separating as well, then:
\begin{equation}
    S_{\Phi \mid \Psi} a|\Phi\rangle= a^{\dagger}|\Psi\rangle
\end{equation}

As we did before, we want to investigate this operator to see what properties it possesses. Analogous to the steps that led to equation (\ref{s conj}), for $S_{\Psi \mid \Phi}$ of an algebra $\mathcal{A}$, we can define  $S_{\Psi\mid\Phi}'$ for the commuting algebra $a' \in \mathcal{A}'$.  Then $S_{\Psi \mid \Phi}' = S_{\Psi \mid \Phi}^\dag$. To prove this we just  show that for all states $\Lambda, \chi$, we have $\left\langle S_{\Psi \mid \Phi}' \Lambda \mid \chi\right\rangle=\left\langle S_{\Psi \mid \Phi} \chi \mid \Lambda\right\rangle$. It is enough to check this for a dense set of states, so we can take $\chi=a \Psi, \Lambda= a' \Psi$. 
    \begin{align}
        \left\langle S_{\Psi \mid \Phi}' a' \Psi \mid a \Psi\right\rangle &=\left\langle a'^\dagger \Phi \mid a \Psi\right\rangle=\left\langle\Phi \mid a' a \Psi\right\rangle =\left\langle\Phi \mid a a' \Psi\right\rangle\\
        &= \left\langle a^\dagger \Phi \mid a' \Psi\right\rangle=\left\langle S_{\Psi \mid \Phi} a \Psi \mid a' \Psi\right\rangle
    \end{align}

\begin{remark} \label{same state red}
    In the case $\Phi = \Psi$, the ``relative" operator is of that between the same system. $S_{\Psi \mid \Psi}$ reduces to $S_{\Psi}$.
\end{remark}
Analogous to the discussion in equation (\ref{polar dec}), we take the polar decomposition of $S_{\Psi \mid \Phi}$ to be:
\begin{equation}
    S_{\Psi \mid \Phi}=J_{\Psi \mid \Phi} \Delta_{\Psi \mid \Phi}^{1 / 2} \quad , \quad  S_{\Psi \mid \Phi}^\dag  =\Delta_{\Psi \mid \Phi}^{1 / 2} J_{\Psi \mid \Phi} 
\end{equation}
Where $J_{\Psi \mid \Phi}$ is the \textit{ relative modular conjugation}, and $\Delta_{\Psi \mid \Phi}$ is the \textit{relative modular operator}. It's straightforward to see that:
\begin{equation}
    \Delta_{\Psi \mid \Phi}=S_{\Psi \mid \Phi}^{\dagger} S_{\Psi \mid \Phi}
\end{equation}
Remark (\ref{same state red}) follows; in the case $\Phi = \Psi$, the relative modular operator $\Delta_{\Psi \mid \Psi}$ reduces to modular operator $\Delta_{\Psi}$.

It would be of great use to relate the modular operator to commutants of algebra $a'\in\mathcal{A}'$, where $a'$ is unitary. We map $\Phi \rightarrow a'\Phi$, so $S_{\Psi \mid \Phi} \rightarrow S_{\Psi \mid a' \Phi} := a'S_{\Psi \mid \Phi}$. Then
\begin{equation}
    \Delta_{\Psi \mid a' \Phi}= S_{\Psi \mid a' \Phi}^\dag S_{\Psi \mid a' \Phi} = S_{\Psi \mid \Phi} (a'^\dag a') S_{\Psi \mid \Phi}= \Delta_{\Psi \mid \Phi}
\end{equation}

\subsection{Tomita–Takesaki theory and Type III factors}

We've mentioned that in physics we mostly deal with hyperfinite algebra $\mathcal{A}$ because the finiteness properties makes them far easier to to work than the non-hyperfinite case, and they're the ones to mostly show up in physics. We start by considering a simple case of algebra $\mathcal{A}\otimes\mathcal{A}'$ acting on $\mathcal{H}=\mathcal{H}_1\otimes\mathcal{H}_2$ \footnote{This factorization is possible according to the separable decomposition theorem. A tensor product structure (TPS) expression is possible if spectrum of the original Hilbert space $\mathcal{H}$ (including multiplicities of eigenvalues) can be written as the sum of two other spectra $\mathcal{H}_1,\mathcal{H}_2$ \cite{PhysRevLett.92.060402, Cotler_2019}.
} \cite{kornyak2021decomposition}.This tensor product Hilbert space could be thought of as bipartite quantum system. The action of the an element $\mathrm{a}$ of the algebra $\mathcal{A}$ on the space $\mathcal{H}$ is as $\mathrm{a} \otimes 1$, where $\mathrm{a}: \mathcal{H}_1 \rightarrow \mathcal{H}_1$. The action of the a commutant  $\mathrm{a}' \in\mathcal{A}'$ on the space $\mathcal{H}$ is as $1 \otimes \mathrm{a}'$, where $\mathrm{a}': \mathcal{H}_2 \rightarrow \mathcal{H}_2$.

A vector $\Psi\in\mathcal{H}$ can be expressed as:
\begin{equation}
    \Psi=\sum_{k=1}^N c_k \psi_k \otimes \psi_k^{\prime}
\end{equation}
Where dim$(\mathcal{H}_1)=$ dim$(\mathcal{H}_2)=N$, $\{\psi_k\}$ are orthornomral basis for $\mathcal{H}_1$, and $\{\psi_{k}'\}$ are orthornomral basis for $\mathcal{H}_2$. We can see that the action of the algebra on the state is:
\begin{equation}
    (\mathrm{a} \otimes 1) \Psi=\sum_{k=1}^n c_k \mathrm{a} \psi_k \otimes \psi_k^{\prime}
\end{equation}
We note that if $\{\psi_k\}$ are complete for $\mathcal{H}_1$ and if $\mathrm{a} \Psi = 0$, then we can conclude that $\mathrm{a}=0$. Which means that $\Psi$ is separating for $\mathcal{A}$. Following a similar reasoning for $\{\psi_{k}'\}$, we conclude that $\Psi$ is cyclic and separating for $\mathcal{A}$ and $\mathcal{A}'$, $\forall C_k \neq 0$.\\

\textbf{Notations:} Proceeding with formulating Tomita-Takesaki theorem, we  rewrite  $\psi_k \rightarrow |k\rangle$, $\psi_k' \rightarrow |k\rangle'$, and $|j\rangle \otimes |k\rangle' \rightarrow |j,k\rangle$. Hence,
\begin{equation}
    \Psi=\sum_{k=1}^N c_k \psi_k \otimes \psi_k^{\prime} = \sum_{k=1}^N c_k |k\rangle |k\rangle' =  \sum_{k=1}^N c_k |k,k\rangle = \sum_{k=1}^N c_k |k\rangle \langle k|'
\end{equation}
Where the last one is in expressed matrix form.

In analogy to what we have done in section(\ref{tomita}), we would like to find a modular operator. We start by defining Tomita operator $S: \mathcal{H} \rightarrow \mathcal{H}$ 

\begin{equation}\label{s on psi}
    S_{\Psi}((\mathrm{a} \otimes 1) \Psi)=\left(\mathrm{a}^{\dagger} \otimes 1\right) \Psi
\end{equation}

Any $\mathrm{a}\in\mathcal{A}$ has the form $\sum \mathrm{a}_{ji} E_{ji}$, where $E_{ji} = |j\rangle \langle i|$. From here we see that $\mathrm{a}^\dag=\sum \overline{\mathrm{a}_{ji}} E_{ij}$. So,
\begin{equation}
    (\mathrm{a} \otimes 1) \Psi = \left( \sum_{ij} \mathrm{a}_{ji} |j\rangle \langle i| \otimes 1\right) \, \sum_{k} c_k |k\rangle \otimes |k\rangle' = \sum_{ijk} \mathrm{a}_{ji} c_k  |j\rangle \langle i|k\rangle \otimes |k\rangle'   = \sum_{ij} \mathrm{a}_{ji} c_i  |j\rangle \otimes |i\rangle'  
\end{equation}
Similarly,
\begin{equation}\label{a dag on psi}
    \left(\mathrm{a}^{\dagger} \otimes 1\right) \Psi = \sum_{ij}  \overline{\mathrm{a}_{ji}} c_j  |i\rangle \otimes |j\rangle' 
\end{equation}
From equation (\ref{s on psi}), we see that:
\begin{equation}\label{a on psi}
    S_{\Psi}((\mathrm{a} \otimes 1) \Psi) = \sum_{ij} \overline{\mathrm{a}_{ji}} \overline{c_i}   S_{\Psi} |j\rangle \otimes |i\rangle' = \left(\mathrm{a}^{\dagger} \otimes 1\right) \Psi 
\end{equation}
Comparing coefficients of $\overline{\mathrm{a}_{ji}}$ in equations (\ref{a dag on psi}) and (\ref{a on psi}) leads us to find that:
\begin{equation} \label{comp s1}
     S_{\Psi} |j,i\rangle = \frac{c_j}{\overline{c_i}}  |i,j\rangle 
\end{equation}
Furthermore, we can derive the action of the adjoint $S_{\Psi}^{\dagger}$ by utilizing the antilinarity of $S_{\Psi}$:
\begin{equation}
    \langle k,l\mid S_{\Psi}^\dag \mid i,j\rangle =  \langle i,j\mid S_{\Psi} \mid k,l\rangle = \langle i,j\mid \frac{c_k}{\overline{c_l}}  |l, k\rangle = \frac{c_j}{\overline{c_i}} \Rightarrow S_{\Psi}^\dag \mid i,j\rangle   = \frac{c_j}{\overline{c_i}} \mid j,i\rangle
\end{equation}
Which means:
\begin{equation}\label{comp s2}
    S_{\Psi}^\dag \mid j,i\rangle   = \frac{c_i}{\overline{c_j}} \mid j,i\rangle
\end{equation}
Comparing equations (\ref{comp s1}) and (\ref{comp s2}), $S_{\Psi} |j,i\rangle = \lambda_{ij} |i,j\rangle $ and $S_{\Psi}^\dag |j,i\rangle = \frac{1}{\bar{\lambda}_{ij}} |i,j\rangle $.

Having worked out both $S_{\Psi}$ and $S_{\Psi}^\dag$, we can see how the modular operator $\Delta_\Psi$ acts:
\begin{equation}
    \Delta_\Psi \mid i,j\rangle =  S_{\Psi}^\dag S_{\Psi} \mid i,j\rangle
\end{equation}
We can see that it acts as follows:

    \begin{align}\label{act delta}
         \Delta_\Psi \mid i,j\rangle &= S_{\Psi}^\dag S_{\Psi} \mid i,j\rangle = S_{\Psi}^\dag \frac{c_i}{\overline{c_j}}  |j,i\rangle = \frac{\overline{c_i}}{c_j}  S_{\Psi}^\dag  |j,i\rangle =\frac{\overline{c_i}}{c_j} \frac{c_i}{\overline{c_j}}\mid i,j\rangle\\
         &= \frac{|c_i|^2}{|c_j|^2}\mid i,j\rangle
    \end{align}

Where in the third equality, we used the antilinear property to move $S_{\Psi}^\dag$ past the eigenvalue by taking the conjugate of the factor. The spectral composition of the modular operator:
\begin{equation}
    \Delta = \sum_{ij} \frac{|c_i|^2}{|c_j|^2} |i,j\rangle \langle i,j|
\end{equation}

Finally, we know that from equation (\ref{polar dec}) that the polar composition of Tomita operator is given by $ S_{\Psi}=J_{\Psi} \Delta_{\Psi}^{1 / 2}$. Furthermore, spectral theory suggests that if we put the operator in a function, its eigenvalue is affected by the same function. Applying this principle on equation (\ref{act delta}):
\begin{equation}
    f(\Delta) |i,j\rangle = f\left(\frac{|c_i|^2}{|c_j|^2}\right)|i,j\rangle \Rightarrow \Delta^{1/2} |i,j\rangle = \frac{|c_i|}{|c_j|}|i,j\rangle
\end{equation}
Also, having the result of equation (\ref{comp s1}) in mind:
\begin{equation}
    S_{\Psi} |i,j\rangle = J_{\Psi} \Delta_{\Psi}^{1 / 2} |i,j\rangle = J_{\Psi}  \frac{|c_i|}{|c_j|}|i,j\rangle \Rightarrow J_{\Psi} |j,i\rangle = \frac{c_j}{c_i}  \frac{|c_i|}{|c_j|} |i,j\rangle  = |i,j\rangle 
\end{equation}
We can define the action of the relative Tomita operator as:
\begin{equation} \label{act rel s}
    S_{\Psi \mid \Phi}((\mathrm{a} \otimes 1) |\Psi\rangle)=\left(\mathrm{a}^{\dagger} \otimes 1\right) |\Phi\rangle
\end{equation}
for a second state $\Phi$:
\begin{equation}
    \Phi=\sum_{\alpha=1}^n d_\alpha \phi_\alpha \otimes \phi_\alpha^{\prime} = \sum_{\alpha=1}^n d_\alpha |\alpha \alpha \rangle
\end{equation}
Replication what we did for $S_{\Psi}$, we let $\mathrm{a} = E_{\alpha i} = |\alpha \rangle\langle i|$. The left hand side of equation (\ref{act rel s}) yields:
\begin{equation}
    S_{\Psi \mid \Phi}\left(\mathrm{a} \otimes 1) |\Psi\rangle = S_{\Psi \mid \Phi}(|\alpha \rangle \langle i| \otimes 1\right)\left( \sum_{k} c_k |k\rangle \otimes |k\rangle'\right)  = S_{\Psi \mid \Phi} c_k |\alpha \rangle \delta_{ik} \otimes |k\rangle' = \overline{c_k} S_{\Psi \mid \Phi}  |\alpha i \rangle
\end{equation}
As for the right hand side:
\begin{equation}
  \left(\mathrm{a}^{\dagger} \otimes 1\right) |\Phi\rangle = d_\alpha |i\alpha \rangle
\end{equation}
equating both side, we get that:
\begin{equation}
    S_{\Psi \mid \Phi}  |\alpha i \rangle = \frac{d_\alpha}{\overline{c_k}}|i\alpha \rangle
\end{equation}
As for the action of the adjoint:
\begin{equation}
    S_{\Psi \mid \Phi}^{\dagger}|i, \alpha\rangle=\frac{d_\alpha}{\bar{c}_i}|\alpha, i\rangle
\end{equation}
and the relative modular operator:
\begin{equation}
    \Delta_{\Psi \mid \Phi}|\alpha, i\rangle=\frac{\left|d_\alpha\right|^2}{\left|c_i\right|^2}|\alpha, i\rangle
\end{equation}
To gain an insight on the action of the relative modular operator, we consider its spectral decomposition:

    \begin{align}
        \Delta_{\Psi \mid \Phi}&=\sum_{\alpha i} \frac{\left|d_\alpha\right|^2}{\left|c_i\right|^2}|\alpha, i\rangle\langle\alpha, i|\\
        &= \sum_{\alpha i} \frac{\left|d_\alpha\right|^2}{\left|c_i\right|^2} |\alpha\rangle\langle\alpha|\otimes |i\rangle'\langle i|'\\
        &= \sum_{\alpha} \frac{\left|d_\alpha\right|^2}{\left|c_i\right|^2} |\alpha\rangle\langle\alpha|\otimes |1\rangle'\langle 1|' + \sum_{\alpha} \frac{\left|d_\alpha\right|^2}{\left|c_i\right|^2} |\alpha\rangle\langle\alpha|\otimes |2\rangle'\langle 2|' + \cdots\\
        &= \left( \sum_{\alpha} \left|d_\alpha\right|^2 |\alpha\rangle\langle\alpha|\right) \otimes \left(\frac{1}{\left|c_i\right|^2} |1\rangle'\langle 1|'  + \frac{1}{\left|c_i\right|^2} |2\rangle'\langle 2|'\cdots \right)\\
        &= \sigma_1 \otimes \rho_2^{-1}
    \end{align}

To make sense of the inverse, recall that for $\rho = \sum \lambda_i E_i \Rightarrow  f(\rho) = \sum f\left(\lambda_i\right) E_i$, where in this case the function is the inverse.
For a state $\Psi$, we define a density matrix $\rho_{12}=|\Psi\rangle \langle\Psi|$. Similarly, for a state $\Phi$, we define a density matrix $\sigma_{12}=|\Phi\rangle \langle\Phi|$. We can find the density on $\mathcal{H}_n$ with respect to a state by tracing out the other density. For example: $\operatorname{Tr}_{\mathcal{H}_2} [\rho_{12}] = \rho_1$, which is the reduced density matrix on $\mathcal{H}_1$ with respect to $\Psi$. The reduced density matrices of $\Psi$ and $\Phi$ are:
\begin{equation}
    \begin{array}{ll}
\rho_1=\sum\left|c_i\right|^2\left|\psi_i\right\rangle\left\langle\psi_i\right| & \sigma_1=\sum\left|d_\alpha\right|^2\left|\phi_\alpha\right\rangle\left\langle \phi_\alpha \right|\\
\rho_2=\sum\left|c_i\right|^2\left|\psi_i^{\prime}\right\rangle\left\langle\psi_i^{\prime}\right| & \sigma_2=\sum\left|d_\alpha\right|^2 \left| \phi_\alpha^{\prime}\right\rangle\left\langle\phi_\alpha^{\prime}\right|
\end{array}
\end{equation}
Notice how the recipe for the modular operator naturally led to density matrices\footnote{When they exist, at least.}, which are essential in defining entropy and discussing states.\\

\textbf{Notation} It is sometimes more intuitive to express states as matrices by transposing on the second system, so we note the \textit{partial transpose} with tilde $\Tilde{\cdot}:\Psi = \sum c_i \psi_i \otimes \psi_i{ }^{\prime} \rightarrow \Tilde{\Psi}= \sum c_i \psi_i \psi_i^{\prime T} \in \mathcal{M}_n$.\\

Given that, $\Psi$ transforms under the action of $\mathrm{a}\in\mathcal{A}$ and $\mathrm{a}'\in\mathcal{A}'$  as:
\begin{equation}
    \begin{aligned}
 (\mathrm{a} \otimes 1) \Psi &\longleftrightarrow && \mathrm{a} \Tilde{\Psi} \\
 \left(1 \otimes \mathrm{a}^{\prime}\right) \Psi &\longleftrightarrow &&\Tilde{\Psi} \mathrm{a}^{\prime T}
\end{aligned}
\end{equation}
We can also see that the trace is:
\begin{equation}
    \langle \Tilde{\Psi} \mid \Tilde{\Phi}\rangle=\operatorname{Tr} \Tilde{\Psi}^{\dagger} \Tilde{\Phi} 
\end{equation}
Now $\Delta_{\Psi \mid \Phi}: \mathcal{M}_n \rightarrow \mathcal{M}_n$. For example, the action of the relative modular operator on the partial transpose of state $\xi$ is:
\begin{equation}
    \Delta_{\Psi \mid \Phi}(\Tilde{\xi}) = \sigma_1 \Tilde{\xi} (\rho_2^{-1})^T = \sigma_1 (\Tilde{\xi}) \rho_1^{-1}
\end{equation}
Given that $\mathcal{H}_1$ is the dual of $\mathcal{H}_1$, which means $\rho_2^{T}=\rho_1$. More generally, 
\begin{equation}
    \Delta_{\Psi \mid \Phi}^\alpha (\Tilde{\xi}) = \sigma_1^\alpha (\Tilde{\xi}) \rho_2^{-\alpha}
\end{equation}
and 
\begin{equation}
    \Delta_{\Psi} (\Tilde{\xi}) = \rho_1 (\Tilde{\xi}) \rho_1^{-1}
\end{equation}
Furthermore, notice that:
\begin{equation}
    \Tilde{\Psi} \Tilde{\Psi}^\dag = \sum c_i \psi_i \psi_i^{\prime T} \,  \overline{\psi}_k' \psi_k^{\dag} \overline{c_k} = \sum c_i \psi_i \psi_i^{\dag} \overline{c_i} = \sum |c_i|^2  \psi_i \psi_i^{\dag} = \rho_1
\end{equation}
Similarly, one can deduce that $ \Tilde{\Psi}^\dag \Tilde{\Psi} =\rho_2$.

Additionally, there are important properties which involve the modular automorphism group, which is the group of unitary transformations $\Delta_{\Psi}^{\mathrm{i} s}, s \in \mathbb{R}$. The modular automorphism group has applications in a variety of areas, particularly it provides a way to understand the entanglement properties of quantum field theory. We can see that this group maps $\mathcal{A}$ to itself. For $\mathrm{a} \otimes 1 \in \mathcal{A}$:
\begin{equation}
    \Delta_{\Psi}^{\mathrm{i} s}(\mathrm{a} \otimes 1) \Delta_{\Psi}^{-\mathrm{i} s}= \rho_1^{\mathrm{i} s} \mathrm{a} \rho_1^{-\mathrm{i} s} \otimes \rho_2^{-\mathrm{i} s} \rho_2^{\mathrm{i} s}=\rho_1^{\mathrm{i} s} \mathrm{a} \rho_1^{-\mathrm{i} s} \otimes 1
\end{equation}
Which can be abbreviated as $\Delta_{\Psi}^{\mathrm{i} s} \mathcal{A} \Delta_{\Psi}^{-\mathrm{i} s}=\mathcal{A}$.
\begin{remark}
    If $\rho$ is thermal (according to Boltzman distribution),then the modular operator corresponds to the Heisenberg dynamics: $\rho_1 \rightarrow e^{-H}$ then $\rho_1^{\mathrm{i} s} \mathrm{a} \rho_1^{-\mathrm{i} s} = e^{-\mathrm{iH}}\mathrm{a} e^{\mathrm{iH}s}$.
\end{remark}
Similarly, for $1 \otimes \mathrm{a}' \in \mathcal{A}'$, $\Delta_{\Psi}^{\mathrm{is}} \mathcal{A}^{\prime} \Delta_{\Psi}^{-\mathrm{i} s}=\mathcal{A}^{\prime}$: 
\begin{equation}
    \Delta_{\Psi}^{\mathrm{i} s}(1 \otimes \mathrm{a}') \Delta_{\Psi}^{-\mathrm{i} s}= 1\otimes \rho_2^{-\mathrm{i} s} \mathrm{a}'\rho_2^{\mathrm{i} s}
\end{equation}
As for the conjugate operator, it maps the algebra to its commutant:
\begin{equation}
    J_{\Psi}(\mathrm{a} \otimes 1) J_{\Psi}=1 \otimes \mathrm{a}^*
\end{equation}
Here $\mathrm{a}^*$ is the conjugate matrix to $\mathrm{a}$. This can be abbreviated as $ J_{\Psi} \mathcal{A} J_{\Psi}=\mathcal{A}^{\prime}$
Similarly, 
\begin{equation}
   J_{\Psi}(\mathrm{a} \otimes 1) J_{\Psi}=1 \otimes \mathrm{a}^*
\end{equation}
Which can be abbreviated as $J_{\Psi} \mathcal{A}^{\prime} J_{\Psi}=\mathcal{A}$.

The \textit{relative modular group} $\Delta_{\Psi \mid \Phi}^{\mathrm{is}}, s \in \mathbb{R}$ acts on the element of the algebra as:
\begin{equation}
    \Delta_{\Psi \mid \Phi}^{\mathrm{i} s}(\mathrm{a} \otimes 1) \Delta_{\Psi \mid \Phi}^{-\mathrm{i} s}=\sigma_1^{\mathrm{i} s} \mathrm{a} \sigma_1^{-\mathrm{i} s} \otimes \rho_1 \rho_1^{-1} = \sigma_1^{\mathrm{i} s} \mathrm{a} \sigma_1^{-\mathrm{i} s} \otimes 1
\end{equation}
Recall that $\sigma$ corresponds to $\Phi$ and $\rho$ corresponds to $\Psi$. This leads to an additional property the ones that the regular modular operator has: the relative modular operator depends on $\Phi$ only and not on $\Psi$. If $\Psi$ and $\Psi'$ are two cyclic separating vectors:
\begin{equation}
    \Delta_{\Psi \mid \Phi}^{\mathrm{i} s}(\mathrm{a} \otimes 1) \Delta_{\Psi \mid \Phi}^{-\mathrm{i} s}=\Delta_{\Psi^{\prime} \mid \Phi}^{\mathrm{is} s}(\mathrm{a} \otimes 1) \Delta_{\Psi^{\prime} \mid \Phi}^{-\mathrm{i} s}
\end{equation}

In conclusion, the Tomita-Takesaki theorem is a powerful result that has far-reaching implications in the study of operator algebras in quantum field theory. This theorem provides a framework for understanding the structure of von Neumann algebras, which are central to the study of infinite-dimensional quantum systems. The theorem shows that any von Neumann algebra is isomorphic to the algebra of bounded operators on a Hilbert space equipped with a certain anti-unitary involution operator, known as the modular conjugation. The modular conjugation plays a crucial role in the theory, as it encodes the spatial symmetry of the system \cite{Rosso_2020}.

Moreover, the Tomita-Takesaki theorem has deep connections to many other areas of mathematics and physics, including the theory of group representations, the theory of automorphic forms, and the study of quantum field theory on curved spacetime. It has also been used to investigate entanglement entropy in quantum field theory \cite{nelson}, which is an important tool for understanding the quantum properties of many-body systems \cite{Dalmonte}.

Overall, the Tomita-Takesaki theorem provides a powerful tool for investigating the structure of operator algebras in quantum field theory, and has broad applications across many areas of physics and mathematics. Its importance cannot be overstated, and it continues to be an active area of research today.

\let\cleardoublepage\clearpage
\part{}

\chapter{Algebraic Quantum Field Theory (AQFT)}\label{4}

In the previous section, we discussed the role of von Neumann algebras in QFT. In this section, we introduce the concept of local operator algebra approach to QFT, which provides a powerful framework for understanding the algebraic structure of the theory.

In the local operator algebra approach, one considers the algebra of observables associated with a subregion of spacetime. The idea is that physical observables can be measured in a finite region of spacetime, and the algebra of these observables is generated by the local observables associated with the subregions of spacetime. This algebra is referred to as a local algebra, and it is a subalgebra of the von Neumann algebra associated with the entire spacetime.

The local operator algebra approach has several advantages. First, it provides a natural framework for understanding the algebraic structure of QFT, including the existence of a vacuum state, the existence of a spectrum condition, and the existence of a Poincaré group action
\cite{witten2018}. Second, it allows one to formulate the notion of locality in a precise way, since the algebra of observables associated with a subregion of spacetime only depends on the observables associated with that region and its causal complement \cite{ULRICH}. Finally, it provides a powerful tool for studying the entanglement structure of QFT, as we will see in this section.

\section{Local Operator Algebraic Approach to Quantum Field Theory}

\begin{figure}[H] \label{fig region}
    \centering\includegraphics[width=9cm, height=6cm]{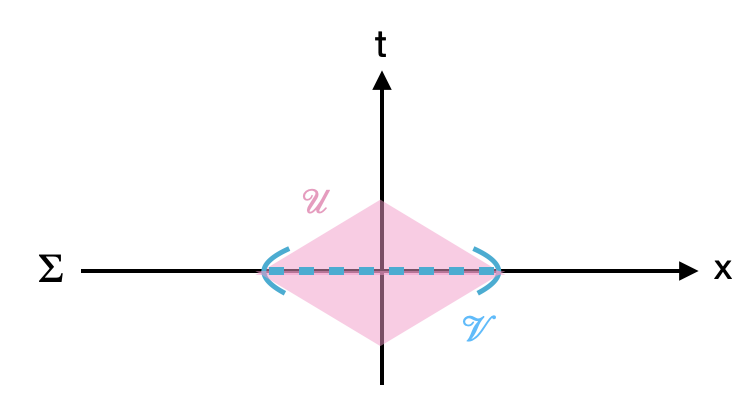}
    \caption{At $t=0$ we define a spacelike hypersurface $\Sigma$, and select a subregion $\mathcal{V}$ in it, then take a local neighborhood $\mathcal{U}$ in spacetime on which $f$ is supported.}
\end{figure}

The local operator algebra approach has its roots in the work of Haag and Kastler \cite{Haag:1963dh}, who showed that the algebra of observables associated with a bounded region of spacetime can be constructed from the algebra of observables associated with the entire spacetime. We begin the discussion by defining a general scalar quantum field $\phi(x)$ that defined on Minkowski spacetime manifold $M=\left(\mathbb{R}^4, \eta\right)$. \textit{Reeh-Schlieder theorem},  states that `any state' in a quantum field theory can be obtained by applying a local operator to the cyclic separating vacuum state \cite{reeh1961bemerkungen,fewster2019algebraic}.   Hence, we define the vacuum state to be $|\Omega\rangle$ which lives in the Hilbert space $\mathcal{K}$. When the field operator acts on the vacuum vector we can write
\begin{equation}
    \phi |\Omega\rangle = |\Phi\rangle,
\end{equation}
Here we notice a problem, the norm of the generated state is infinite $\langle\Phi\mid\Phi\rangle=\infty$, because spacetime itself is taken to be infinite and non-compact. To get around this, we define a spacelike Cauchy hypersurface $\Sigma$. On that hypersurface, we choose a finite open region $\mathcal{V} \subset \Sigma$. % that is still spacelike. 
Because $\Sigma$, and thus $\mathcal{V}$, is like a point with respect to the time-like direction orthogonal to $\Sigma$. A point is not open, so we'd like to consider an open neighborhood of the spacetime, $\mathcal{U}$, around $\mathcal{V}$. 
%Lastly, we take a small deviation $\mathcal{U}$ around $\mathcal{V}$.
Now we have a local region of spacetime called $\mathcal{U}_\mathcal{V}$, as shown in figure (\ref{fig region}), and we are set to fix the divergence problem \cite{klaus}.

We define a smooth compact \textit{smearing function} $f$ that is supported  on such a neighborhood $\mathcal{U}$. We use \textit{smeared field operator} $\phi_f = \int \mathrm{d}^4 x f(x) \phi(x) \in \mathcal{A}_\mathcal{U}$ to generate the states:
\begin{equation}
    \phi_{f_1}\phi_{f_2} \cdots \phi_{f_n}|\Omega\rangle = |\Phi_{\Vec{f}} \rangle
\end{equation}
These states norm of this state convergent $\langle\Phi_{\Vec{f}}\mid\Phi_{\Vec{f}}\rangle <\infty$, which means it's indeed a vector in $\mathcal{K}$. 
In QFT, the type of von Neumann algebra associated with a spacetime region depends on the causal structure of that region. Specifically, if a region is causally disconnected from another region, then the corresponding von Neumann algebras are type I factors. If two regions are causally connected but not in a time-like or light-like relationship, then the corresponding von Neumann algebras are type II factors. Finally, if two regions are in a time-like or light-like relationship, then the corresponding von Neumann algebras are type III factors.
The fact that different types of von Neumann algebras are associated with different regions of spacetime is important because it suggests that different regions of spacetime have different quantum mechanical properties. For example, two regions that are causally disconnected from each other might have completely different sets of observables, even if they have the same underlying physical system. This highlights the non-locality of QFT and the importance of operator algebras in characterizing its structure.

\subsection{Haag duality}

Haag duality states that for a certain class of quantum field theories, called local field theories, there is a duality between the algebra of a region of spacetime and the algebra of its complement \cite{haag}. Given a local field theory and a region $\mathcal{U}$ of spacetime, the algebra $\mathcal{A_U}$ associated with that region is isomorphic to the algebra $\mathcal{A_{U}}$ of associated with its complement region $\mathcal{U}'$, where the regions are space-like separated. This means that all physical information that can be obtained by measuring observables in region $\mathcal{U}$ can also be obtained by measuring observables in its complement $\mathcal{U}'$.

\begin{figure}[H]
    \centering\includegraphics[width=9cm, height=6cm]{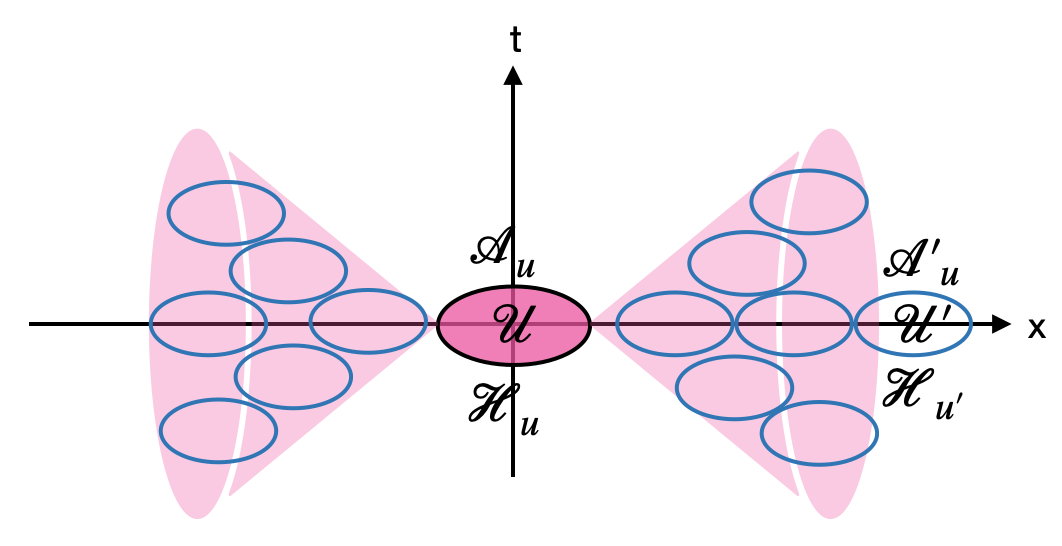}
    \caption{For a region $\mathcal{U}$ with an associated algebra $\mathcal{A_U}$, there's a complement space-like separated region $\mathcal{U}'$ with commutant algebra $\mathcal{A_U}'$}\label{fig:my_label}
\end{figure}

To see this, consider the vacuum $|\Omega\rangle \in \mathcal{K}$ that is cyclic for $\mathcal{A}'$. Let  $\mathrm{a}\in\mathcal{K}$. If $\mathrm{a}|\Omega\rangle=0$, then $\mathrm{a}\mathcal{K}=0$ if and only if $\mathrm{a}=0$:
\begin{equation}
    \mathrm{a}\mathcal{K}= \mathrm{a} \mathcal{A}' |\Omega\rangle=\mathcal{A}' \mathrm{a}  |\Omega\rangle = 0
\end{equation}
which means that the vacuum must be separating for $\mathcal{A}$. The vacuum separating for the algebra if and only if it is cyclic for its commutant, and vice-versa.

In essence, Haag duality connecting space-time subsets to von Neumann algebra, simply put: complements $\longleftrightarrow$ commutants \cite{Longo_2020}. If we have two complement spaces, their algebra are each other’s commutants. This can be expressed as:
\begin{equation}
\mathcal{A}_{\mathcal{U}^{\prime}}=\mathcal{A}_{\mathcal{U}}^{\prime}
\end{equation}

While the Haag duality has been proven in various settings, there are also known counterexamples which suggest that the principle may not hold in all cases \cite{GARBARZ2022115797,dedushenko2023snowmass}. Here are some examples of counterexamples to the Haag duality:
\begin{itemize}
    \item One such case is when there is a nontrivial topology of the underlying spacetime. In this case, the causal complement of a region may not be well-defined, and it may not be possible to express all observables in terms of observables outside the region \cite{witten1989quantum}.
    \item Another case where Haag duality may fail is when the system under consideration is not in a pure state. In this case, the algebra of observables may not be maximal, and it may not be possible to express all observables in terms of the observables in the causal complement of a region \cite{pittphilsci649}.
    \item A third case is the Doplicher–Haag–Roberts (DHR) sector: The DHR sector is a class of representations of the algebra of observables in a relativistic quantum field theory that are associated with localized particles. In some theories, such as those with long-range forces, there can be DHR sectors that are not generated by the observables localized in bounded regions of spacetime \cite{doplicher1974local}. 
\end{itemize}
It is worth noting that despite these counterexamples, the Haag duality is still considered to be a fundamental principle in AQFT, as it holds in many important cases and is a key tool in understanding the algebraic structure of quantum field theories.

\section{Entanglement Entropy} \label{ent entropy}

Entanglement and entropy are related through the concept of \textit{entanglement entropy}, which measures the amount of entanglement or shared information between subsystems in a larger quantum system \cite{headrick2019lectures}. In this context, entanglement entropy provides a way to characterize the long-range correlations between different parts of the system. To see the relation between entropy and entanglement, let's consider the worked example (\ref{rho exmp}):
    \begin{align}
        \varrho & = |\Psi \rangle \langle \Psi| \\
        &= \left(\sum_{i} c_i | \psi_i \rangle \otimes | \phi_i \rangle \right)\left(\sum_{j}\Bar{c_j}\langle \psi_j|\otimes \langle\phi_j |\right)= \sum_{i,j} c_i \Bar{c_j}(| \psi_i \rangle \langle \psi_j| \otimes | \phi_i \rangle \langle \phi_j |)
    \end{align} 
    
We can consider the density matrix $\varrho$ to be a joint state of $A\otimes B$, where $A=| \psi_i \rangle \langle \psi_j|$ and $B = | \phi_i \rangle \langle \phi_j |$. A natural question here is \textit{what's the state on $A$?} or \textit{how entangled is $A$?}  We can answer this by taking the trace of the density matrix with respect to $B$:
    \begin{align}\operatorname{Tr}_B[\varrho] & = \operatorname{Tr}_B[|\Psi \rangle \langle \Psi|] = \operatorname{Tr}_B[A \otimes B]=A \operatorname{Tr}[B]\\
         & = c_1 \Bar{c_1}(| \psi_1 \rangle \langle \psi_1|\otimes | 1) + c_2 \Bar{c_2}(| \psi_2 \rangle \langle \psi_2|\otimes 1)\\
         &+ c_1 \Bar{c_2}(| \psi_1 \rangle \langle \psi_2| \otimes 0) + c_2 \Bar{c_1}(| \psi_2 \rangle \langle \psi_1|\otimes 0)\\
         &= |c_1|^2 \begin{pmatrix}
            1 & 0 \\
            0 & 0\\
            \end{pmatrix}  + |c_2|^2 \begin{pmatrix}
            0 & 0 \\
            0 & 1\\
            \end{pmatrix}\\
            &= \begin{pmatrix}
            p_1 & 0 \\
            0 & p_2\\
            \end{pmatrix}
    \end{align} 

Where the value of $p_1$ and $p_2$ determine the entropy or mixture of the $\Psi$. For example, if $p_1=1 , p_2 =0$, we have a pure state, and if $p_1=1/2 , p_2 =1/2$, we have maximum mixture (or maximally entangled) state. Furthermore, if $\operatorname{Tr}[\varrho] = pure \Rightarrow \varrho$ is NOT entangled, and if it's \textit{mixed}, then $\varrho$ is entangled. So we see a direct correlation between entanglement and entropy.

We can describe entanglement as a  tensor product between system. Most basic case is that of 2-qubits: $\mathbb{C}^2 \otimes \mathbb{C}^2 = \mathbb{C}^4 = \mathcal{H}$. Take $\psi, \phi \in \mathbb{C}^2$. $\psi \otimes \phi$ define a pure state $\omega: \mathcal{A} \rightarrow \mathbb{C}$:
\begin{equation}
    \omega (x) = \operatorname{Tr} [\varrho x] = \langle \psi \otimes \phi \mid x \mid \psi \otimes \phi \rangle
\end{equation}
Where the density matrix $\varrho = | \psi \rangle \langle \phi |$ and $x \in \mathcal{B(H)}$.

There are multiple definitions of entropy, one of the most common is  von Neumann entropy:
\begin{equation}
    S = -\operatorname{Tr}[\varrho \ln{\varrho}]
\end{equation}
However, most definitions of entropy rely on the existence of a trace. In the case of infinite degrees of freedom, such as in QFT, a notion of trace does not exist, so we can't use the standard forms of entropy. This is because the density matrices are von Neumann algebras of Type $III$, not of Type $I$. Given the Tomita-Takesaki theorem, we can generalize the standard formalism of entropy to the relative entropy $\mathcal{S}_{\Psi \mid \Phi}(\mathcal{U})$  \cite{araki1976relative} between two states $\Phi$ and $\Psi$ in local region $\mathcal{U}$ of spacetime:
\begin{equation}
\mathcal{S}_{\Psi \mid \Phi}(\mathcal{U})=-\left\langle\Psi\left|\log \Delta_{\Psi \mid \Phi}\right| \Psi\right\rangle
\end{equation}
The Tomita-Takesaki theorem is intimately related to entanglement entropy in QFT. The theorem provides a way to construct a unique modular conjugation operator, which in turn can be used to define a unique vacuum state \cite{Ipek_2019}. This vacuum state is important for the calculation of entanglement entropy in QFT, as it serves as a reference state against which the entanglement of other states can be measured. The span of the relative entropy $\mathcal{S}_{\Psi \mid \Phi}(\mathcal{U}) \in [0,\infty)$, so we know that Type $III_1$ factors are suitable to describe QFT. More precisely, it only vanishes when $\Phi=\mathrm{a}^{\prime} \Psi$ for $\mathrm{a}' \in\mathcal{A}_{\mathcal{U}'}$ being unitary. If that's the case, then by equation ($\Delta_{\Psi \mid \Phi} \rightarrow \Delta_{\Psi}$), and: 
\begin{equation}
    S_{\Psi \mid a^{\prime} \Psi}=-\left\langle\Psi\left|\log \Delta_{\Psi}\right| \Psi\right\rangle = 0
\end{equation}
because $\log(\Delta_{\Psi})\Psi\rangle = \log({1})|\Psi\rangle =0$, which follows from $f(\Delta_ \Psi)|\Psi\rangle=f(1)|\Psi\rangle$.

 In AQFT, the observables of the theory are represented by operator algebras acting on a Hilbert space. By studying the entanglement between subsystems, one can gain insight into the algebraic structure of the theory and the nature of the correlations between the observables.

Furthermore, entanglement entropy provides a useful tool for understanding the behavior of quantum field theories on curved spacetimes. In such cases, the algebraic structure of the theory can become more complicated due to the presence of non-trivial spacetime geometries. However, by studying the entanglement entropy between subsystems, one can still gain insights into the underlying algebraic structure of the theory. We will further investigate this in the next chapter when we analyze everything discussed in Part I of the essay in the scope of quantum field theory.

In these cases, the quantum dynamics cannot be described by operators acting on a Hilbert space, but instead, must be described by the evolution of density matrices or their generalizations, known as Type III density matrices.

\chapter{Applications}\label{5}
In this section, we explore the applications of the operator algebra approach and entanglement entropy to QFT on spacetime. QFT on curved spacetime is a subject of great interest in theoretical physics, as it provides a way to study the interplay between gravity and quantum mechanics.

One important result in this context is the Unruh effect, which predicts that a uniformly accelerating observer will detect a thermal bath of particles where an inertial observer would detect none. The Unruh effect is a consequence of the fact that the vacuum state of QFT on curved spacetime is observer-dependent \cite{Fedotov_1999}.

Another important application of the operator algebra approach to QFT on curved spacetime is the study of black hole entropy \cite{Chandrasekaran:2022cip}. Black holes are believed to be described by a classical solution to the Einstein field equations, but the entropy associated with a black hole is a purely quantum mechanical concept \cite{Hawking1975ParticleCB}. In this chapter, investigate one of these applications, and we follow a similar approach to Witten's \cite{witten2022does}.

\section{ Quantum field theory on curved spacetime}

We always assume that the spacetime we consider is globally hyperbolic. This means that the spacetime has a complete Cauchy hypersurface $\Sigma$ on which initial conditions for classical or quantum fields can be formulated. The manner in which quantum field theory can be formulated on such a spacetime depends heavily on whether $\Sigma$ is compact. If the hypersurface is compact, we can use the usual Hilbert space formulation of quantum field theory. We can naturally associate a Hilbert space $\mathcal{H}$ with the given theory, and the quantum dynamics can be described in terms of operators acting on $\mathcal{H}$. However, unlike in the case of quantum field theory in Minkowski spacetime, there is usually no distinguished ground state or vacuum vector in $\mathcal{H}$, since there is no natural ``energy" to minimize. That means different states correspond to different observes, and they don’t necessarily agree on the same vacuum. Hence, while formulating quantum field theory in curved spacetime, we must accept the idea of a naturally defined Hilbert space that does not have any distinguished vector.

In the context of quantum field theory in curved spacetime, we explore the challenges of applying Euclidean field theory techniques to Lorentz signature $(-,+,+,...,+)$  on D-dimentional spacetimes manifold. The primary issue is that a generic Lorentz signature spacetime lacks a useful Euclidean continuation. We start by formulating the space and algebra of the quantum field theory. At the end,  we draw parallels with the algebra that we already built in previous sections. We find that the algebra of the theory is equivalent to Type $II$ and Type $III$.

We begin by defining a general quantum scalar field $\phi(\vec{x}, t)$ of mass m. The action of the field can be defined as:
\begin{equation}
    I=-\frac{1}{2} \int \mathrm{d}^D x \sqrt{g}\left(g^{\mu \nu} \partial_\mu \phi \partial_\nu \phi+m^2 \phi^2\right)
\end{equation}
The mode (oscillator) expansion of the field (in 4-D) in terms of creation and annihilation operators $(a_k , a_k^{\dagger})$ and their corresponding mode functions of positive and negative frequency $(f_k(\vec{x}),\bar{f}_k(\vec{x}))$ is given by:
\begin{equation}
    \phi(\vec{x}, t)=\sum_k\left(a_k f_k(\vec{x}) e^{-\mathrm{i} \omega_k t}+a_k^{\dagger} \bar{f}_k(\vec{x}) e^{\mathrm{i} \omega_k t}\right)
\end{equation}
The sum runs over all modes $k$, with $\omega_k$ being the positive frequency associated with each mode. The modes are each given by the canonical variables $\left(\hat{x}_k, \hat{p}_k\right)$\footnote{Commutation relations are as in equation (\ref{comm}).} on an infinite dimensional Hilbert space $\mathcal{H}_k$. A Hamiltonian in the algebra $\mathcal{A}$ which could act on said space is defined as:
\begin{equation}
     H_k =\frac{1}{2}\left(\hat{p}_k^2+\omega_k^2\hat{x}_k^2\right) \equiv \omega_k a_k^{\dag} a_k
\end{equation}

\begin{figure}[H] \label{projection}
    \centering\includegraphics[width=4cm, height=4.5cm]{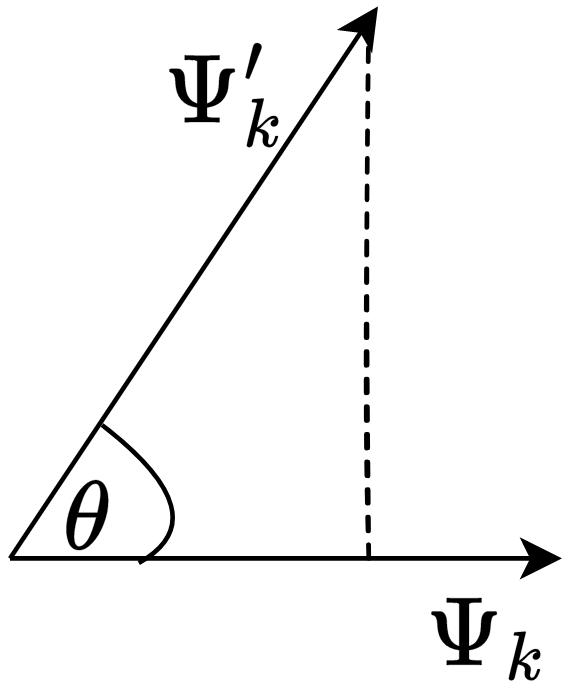}
    \caption{The projection of $\psi_k^{\prime}$ onto $\psi_k$ is defined as $\left|\left\langle\psi_k^{\prime} | \psi_k\right)\right| = \left|\cos(\theta)\right|$.}
\end{figure}

With the commutation relation $\left[a_k, a_l\right]=[a_k^{\dagger}, a_l^{\dagger}]=0,[a_k, a_l^{\dagger}]=\delta_{k l}$. $\mathcal{H}_k$ represents a single space, we can construct an uncountably infinite-dimensional Hilbert space $\mathcal{H^*}$ that encompasses all possible states in each $\mathcal{H}_k$. Then we have:
\begin{equation}
    \mathcal{H^*}=\bigotimes_{k=1}^{\infty} \mathcal{H}_k
\end{equation}
We’re interested on how the algebra acts on states in the space. For a normalized\footnote{Normalization condition: $\left\langle\psi_u \mid \psi_k\right\rangle=1 \Rightarrow\langle\Psi \mid \Psi\rangle=\Pi_k\left\langle\psi_k \mid \psi_k\right\rangle$.} state $\Psi=\bigotimes_{k} \psi_k$ that's generating $\mathcal{H}_\Psi \subset \mathcal{H^*}$, we can generate the space $\mathcal{H}_\Psi$ by taking the completion of action of an algebra on the vector: $\mathcal{A}_0 \Psi$, where $\mathcal{A}_0$ is the algebra all polynomials in the $\hat{p}_k$’s and $\hat{x}_k$’s. Also consider a different state $\Psi'=\bigotimes_{k} \psi'_k$ generating $\mathcal{H}_{\Psi'}\subset \mathcal{H^*}$. A very profound question we could ask is: \emph{Are $\mathcal{H}_{\Psi}$ and $\mathcal{H}_{\Psi'}$ the same space?} The answer to this will give us an insight into the uniqueness of the irreducible representation. We can start off by considering the projection: $\left|\langle\psi_k^{\prime} |\psi_k\rangle\right| \:=c_k$, where $0 \leq c_k \leq 1$, as illustrated in diagram (\ref{projection}).
So $\left|\left\langle\Psi^{\prime} \mid \Psi\right\rangle\right|=\prod_{k=1}^{\infty} c_k=0$ unless $c_k \rightarrow 1$ as $k \rightarrow \infty$. Which means $c_k$ would converge to 1 because only finitely many don’t go to 1, notice that this is very similar to Type $II$ and $III$. In essence, if $\left|\left\langle\Psi^{\prime} \mid \Psi\right\rangle\right| \neq 0$ then $\psi_k^{\prime}\cong \psi_k$ (coincide) $\forall k \geqslant \kappa$, where  $\kappa$ is called the ultraviolet cut-off. Without loss of generality, the projection has two cases:
\begin{enumerate}[i]
\item  $\left|\left\langle\Psi^{\prime} \mid \Psi\right\rangle\right| \neq 0$\\
\item $\left|\left\langle\Psi^{\prime} \mid \Psi\right\rangle\right|=0$
\end{enumerate}
It only takes a single $c_k$ to be zero\footnote{which correspond vectors being orthogonal with $\theta=90^\circ$.}  for the projection to be zero as well. So we remove any $c_k =0$ from the product $\prod_{k=1}^{\infty} c_k$, then re-evaluate the product as $k \rightarrow \infty$. After getting rid of $c_k =0$, we look at the product:
\begin{equation}
    \prod_{k=1}^{\infty}  c_k=\left\{\begin{array}{l}
\neq 0 \quad \quad \quad c_k \rightarrow 1 \quad \text{rapidly as} \quad k \rightarrow \infty\\
  0 \quad \quad \quad \quad c_k \nrightarrow 1 \quad \text{as} \quad k \rightarrow \infty
\end{array}\right.
\end{equation}
We proceed with analyzing case (i):
We find that $\Psi$ and $\Psi'$ are different (don't coincide) for \textit{finitely} many $\Psi_k$'s. We can use $\mathcal{A}_0$ to see that $\mathcal{H}_{\Psi}=\mathcal{H}_{\Psi'}$.

Analyzing case (ii):
If the projection still yields zero for $c_k \neq 0$, we find that $\Psi$ and $\Psi'$ are different for \textit{infinitely} many $\Psi_k$'s. $\mathcal{A}_0\Psi$ just makes finitely many changes to $\Psi$. So  $\left\langle\Psi^{\prime} \mid \Psi\right\rangle = 0$, but also for $\mathrm{a}\in\mathcal{A}_0$, we that $\left\langle \mathrm{a}\Psi^{\prime} \mid \Psi\right\rangle =0$ too. This applies for the whole algebra as well: $\left\langle \mathcal{A}_0 \Psi^{\prime} \mid \Psi\right\rangle =0$, which means $\left\langle \mathcal{A}_0 \Psi^{\prime} \mid \mathcal{A}_0 \Psi\right\rangle =0$ This means that the spaces are orthogonal:
\begin{equation}
    \mathcal{H}_{\Psi} \perp \mathcal{H}_{\Psi'}
\end{equation}
In this case, the spaces are NOT unique because the algebra generates completely different spaces. They’re not unitarily equivalent\footnote{Uniqueness would imply: $\mathcal{H}_{\Psi} =U \mathcal{H}_{\Psi'} \cong \mathcal{H}_{\Psi'}$, where $U$ is unitary.}.  We cannot go from one space to the other via unitary transformation, as we were doing with $\hat{x}$ and $\hat{p}$ in finite degree of freedom (QM). This goes in-line to what Stone-von Neumann theorem told us, that in infinite degrees of freedom, there can be multiple inequivalent irreducible representations.
We can see that we keep doing this with an unlimited choice of different $\Psi$'s, so we can keep reducing $\mathcal{A}_0$ indefinitely, i.e. there's no irreducible representation, as expected in a Type $II$ and $III$ factors.

One consequence of this richer algebraic structure is that it is not possible to define a unique vacuum state in QFT. Instead, there are infinitely many vacua, each associated with a different representation of the canonical commutation relations. This has important consequences for the understanding of particle creation and annihilation in QFT.
So in `large' space $\mathcal{H^*}=\bigotimes_{k=1}^{\infty} \mathcal{H}_k$, we have the subspace $ \mathcal{H}_k \supset \mathcal{A}_0 \Psi$. For $\Psi \neq \Psi'$, we have $\mathcal{H}_{\Psi} = \mathcal{H}_{\Psi'}$ iff $\Psi|_{k \geqslant \kappa} = \Psi^{\prime}|_{k \geqslant \kappa}$ (at high energy, states are the same). Which means that at low energy modes, where $k < \kappa$, the choice of $\psi_k$ is arbitrary for finitely many $k$'s.

For a local region $\mathcal{U}\subset M$, consider the corresponding algebra $\mathcal{A_U}$ which obey the physically motivated axioms 
\cite{Hollands_2009}. For example, \textit{causality} implies that $\mathcal{A_U}$ commutes with $\mathcal{A_{U'}}$ if $\mathcal{U}$ and $\mathcal{U'}$ are spacelike separated. Important to note here that because $\mathcal{A_U}$ is von Neumann algebra of Type $III$, it does NOT have an irreducible representation in a Hilbert space. As a consequence, there is no Hilbert space $\mathcal{H_U}$ associated with $\mathcal{U}$. This is because the Reeh-Schlieder theorem indicates $\mathcal{A_U} \Psi$ is dense in $\mathcal{H}$ that's not restricted to $\mathcal{U}$, so $\mathcal{H} \neq \bigotimes_{\cup\mathcal{W}_i} \mathcal{H}_i$. Different region, say $\mathcal{W}$, gives the same space $\mathcal{H}$. If the $\mathcal{A_U}$ was Type $I$, then it would have an irreducible representation in the Hilbert space, and we could obtain a \textit{local Hilbert space} $\mathcal{H_U}$. Despite not being able to find $\mathcal{H_U}$ to identify with $\mathcal{U}$, we can still identify $\mathcal{A_U}$ with $\mathcal{U}$. This means that we have \textit{local algebra}, but not local hilbert space.

To summarize this section, in standard quantum mechanics, the state of a system is described by a vector in a Hilbert space, and the evolution of the state is given by a unitary transformation acting on the Hilbert space. However, in certain cases, such as when the system is subject to strong interactions or in the presence of a gravitational field, the Hilbert space may not be unique or may not exist at all. This is because the concept of a Hilbert space requires the existence of a scalar product, which may not be well-defined in these situations\cite{Ciolli:2021otw}.

In such cases, the quantum state of the system is described by a density matrix, which is a more general mathematical object that can describe mixed states as well as pure states. The Type $III$ generalization of density matrices allows for even more general descriptions of the quantum state, which can be used in situations where the Hilbert space is not well-defined. In situations where the Hilbert space cannot be uniquely defined, the dynamics of quantum systems are described through the evolution of density matrices or, more specifically, Type $III$ generalization of density matrices, rather than through operators on a Hilbert space. This is because a natural energy cannot be minimized and there is no distinguished ground state or vacuum vector in such cases. Therefore, the evolution of the system must be described through the evolution of density matrices.

Overall, the importance of the Type $III$ generalization of density matrices and the Tomita-Takesaki theorem lies in their ability to provide a more general framework for describing quantum systems in situations where the standard Hilbert space formalism is not applicable.

\newpage
\section{Conclusion}
The algebraic approach to QFT provides a powerful framework for understanding the algebraic properties of the observables in a system, particularly in terms of von Neumann algebras. Entanglement entropy provides a measure of the entagnlement.

Moreover, the operator algebra approach to QFT is particularly useful for exploring the phenomenon of entanglement entropy. Entanglement entropy is a measure of the entanglement between different regions of a quantum system. It is defined as the von Neumann entropy of the reduced density matrix of a subsystem, which describes the entanglement between the subsystem and its environment. In QFT, the algebra of observables plays a crucial role in the definition of entanglement entropy.

One of the key insights from the operator algebra approach to QFT is the connection between entanglement entropy and the algebraic structure of QFT. In particular, the concept of local observables is central to understanding the entanglement entropy of a quantum field theory. Local observables are operators that are localized in spacetime regions and can be measured independently of the rest of the system. The algebra of local observables, known as the local operator algebra, contains important information about the entanglement structure of a quantum field theory.

The local operator algebra approach to QFT provides a powerful tool for understanding the entanglement entropy of a quantum field theory. By characterizing the algebraic structure of QFT in terms of local observables, we can compute the entanglement entropy of a subsystem in terms of the algebra of observables associated with that subsystem. This provides a powerful way of characterizing the entanglement structure of a quantum field theory and has important implications for many areas of physics, including quantum gravity and black hole physics.

One example of the application of the operator algebra approach to QFT is in the study of quantum field theory on curved spacetimes. In curved spacetimes, the algebra of observables is no longer trivial and can contain nontrivial topological and geometrical information. The local operator algebra approach provides a powerful tool for understanding the algebraic structure of QFT on curved spacetimes and has important implications for understanding the quantum properties of black holes and providing more insight into some of the highly sought-after aspirations of theoretical physicists \cite{deboer2022frontiers}.
%%%%%%%%%%%%%%%%%%%%%%%%
%%%%%%%%%%%%%%%%%%%%%%%%%%%%%%%%%%%%%%%%%%%%%%%%%%%%%%%%
%% Bibliography:
%%

\setlength{\bibsep}{0.03em} % set the space between bibliography items
\addcontentsline{toc}{chapter}{Bibliography}

\bibliography{thesis}

\newpage
%%%%%%%%%%%%%%%%%%%%%%%%%%%%%%%%%%%%%%%%%%%%%%%%%%%%%%%%%%%%%%%%%%%%%%%%%%%%%%%%
%% Appendix:
%%
\appendix

\end{document}